\documentclass[submit]{smj}
\usepackage{amsthm,amsmath,amsfonts,amssymb}
\usepackage{graphicx, enumerate, multirow, microtype}
\usepackage{booktabs, subfig, bm, paralist, longtable, pdflscape}
\usepackage[normalem]{ulem}

\newcommand{\X}{\mathcal{X}}
\newcommand{\Y}{\mathcal{Y}}
\newcommand{\I}{\mathcal{I}}

\newcommand{\Rlogo}{\protect\includegraphics[height=1.8ex,keepaspectratio]{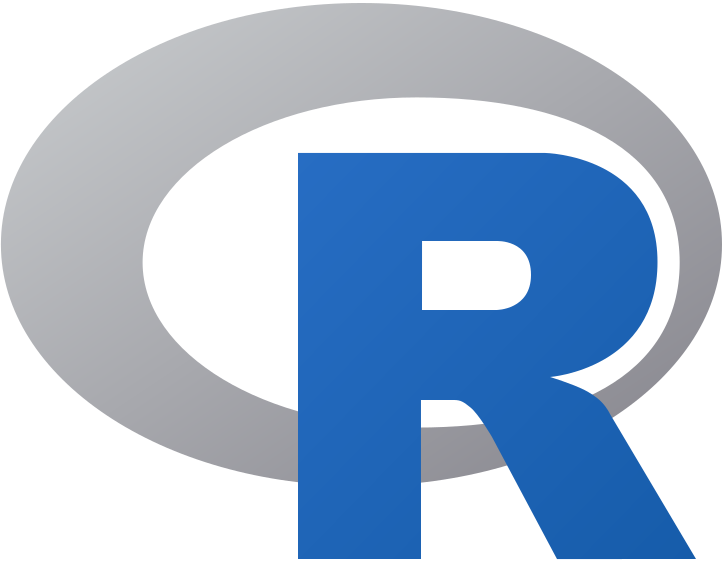}}
\newcommand{\argmax}{\operatornamewithlimits{argmax}}

\DeclareMathOperator*{\argmin}{\arg\!\min}

\theoremstyle{plain}

\theoremstyle{remark}

\newsavebox\CBox

\graphicspath{{plots/}}

\Author{Ufuk Beyaztas\Affil{1}, 
        Han Lin Shang\Affil{2}, 
        and Abhijit Mandal\Affil{3}
}
\AuthorRunning{Ufuk Beyaztas \textrm{et al.}}
\Affiliations
{
\item Department of Statistics, 
      Faculty of Science, 
      Marmara University,
      Istanbul, Turkey

\item Department of Actuarial Studies and Business Analytics,
      Macquarie University, 
      Sydney, Australia

\item Department of Mathematical Sciences,
      University of Texas at El Paso,
      El Paso, USA 
} 

\CorrAddress{Ufuk Beyaztas, 
             Department of Statistics, 
             Faculty of Science,
             Marmara University, 
             34722, Kadikoy-Istanbul, 
             Turkey}
\CorrEmail{ufuk.beyaztas@marmara.edu.tr}
\CorrPhone{(+90)\;216\;777\;3327}
\CorrFax{(+90)\;216\;777\;3201}

\Title{Robust function-on-function interaction regression}
\TitleRunning{Robust function-on-function interaction regression}

\Abstract{
A function-on-function regression model with quadratic and interaction effects of the covariates provides a more flexible model. Despite several attempts to estimate the model’s parameters, almost all existing estimation strategies are non-robust against outliers. Outliers in the quadratic and interaction effects may deteriorate the model structure more severely than their effects in the main effect. We propose a robust estimation strategy based on the robust functional principal component decomposition of the function-valued variables and $\tau$-estimator. The performance of the proposed method relies on the truncation parameters in the robust functional principal component decomposition of the function-valued variables. A robust Bayesian information criterion is used to determine the optimum truncation constants. A forward stepwise variable selection procedure is employed to determine relevant main, quadratic, and interaction effects to address a possible model misspecification. The finite-sample performance of the proposed method is investigated via a series of Monte-Carlo experiments. The proposed method’s asymptotic consistency and influence function are also studied in the supplement, and its empirical performance is further investigated using a U.S. COVID-19 dataset.
}


\begin{document}

\maketitle

\section{Introduction} \label{sec:intro}

Function-on-function regression model (FoFRM), which was first proposed by \cite{ramsay1991}, has become a popular tool for investigating the association between the functional response and one or more functional predictors \citep[see, e.g.,][and references therein]{yao2005, MullerYao2008, wang2014,  ivanescu2015, chiou2016, Greven2017, BS20, BS20C, LuoQi21}. Let $ \left\lbrace \Y_i(t), \bm{\X}_{ip}(s): i = 1, \ldots, n,~ p = 1, \ldots, P \right\rbrace$ denote a random sample from the pair $(\Y, \bm{\X})$, where $\Y \in \mathcal{L}_2(\I)$ and $\bm{\X} = [\X_1, \ldots, \X_P]^\top$ with $\X_p \in \mathcal{L}_2(\mathcal{S})$. Herein, $\Y$ and $\bm{\X}$ respectively denote univariate and $P$-dimensional stochastic processes with elements represented by curves belonging to $\mathcal{L}_2$ Hilbert space defined on the closed and bounded intervals $t \in \I$ and $s \in \mathcal{S}$, where $t$ and $s$ are the observations in the intervals $\I$ and $\mathcal{S}$, respectively. Without loss of generality, we assume that $\I = \mathcal{S} = [0,1]$. Then, the FoFRM is of the following form:
\begin{equation}\label{eq:fof}
\Y_i(t) = \beta_0(t) + \sum_{p=1}^P \int_0^1 \X_{ip}(s) \beta_p(s,t) ds + \epsilon_i(t),
\end{equation}
where $\beta_0(t)$ is the intercept function; $\beta_p(s,t)$ links the functional response with $p$\textsuperscript{th} functional predictor, and it is the $p$\textsuperscript{th} coefficient surface; and $\epsilon_i(t)$ is independent and identically distributed (i.i.d.) random error function with mean zero, and it is independent of $\X_p(s)$ for $ p = 1, 2, \ldots, P$. The FoFRM has become an important tool in functional data analysis literature. Several interesting studies have been proposed in FoFRM, see, e.g.,  \cite{Cai2020}, \cite{Harjit}, and \cite{BS2022BJPS} for robust estimation, and \cite{Wang2022} and \cite{Cai2022} for model selection.

Model~\eqref{eq:fof} considers only the main effects of functional predictors on the functional response. In this model, the relationship between a functional predictor and the functional response is assumed to be independent of other functional predictors. However, this assumption may be restrictive in empirical applications since the functional predictors might interact. Several FoFRM with quadratic and interaction effects of the predictors have been proposed to remedy this problem.

In the context of the FoFRM, \cite{Matsui2020} proposed a simple model with the quadratic effect of the functional predictor as follows:
\begin{equation}\label{eq:sfofqr}
\Y_i(t) = \beta_0(t) + \int_0^1 \X_i(s) \beta(s,t) ds + \int_0^1 \int_0^1 \X_i(r) \X_i(s) \gamma(r,s,t) dr ds + \epsilon_i(t),
\end{equation}
where $\gamma(r,s,t)$ denotes the coefficient hypersurface for the quadratic term $\X_i(r) \X_i(s)$. For Model~\eqref{eq:sfofqr}, \cite{Matsui2020} used a basis function expansion to approximate the two- and three-dimensional model parameters. In addition, \cite{SunWang} proposed a functional principal component analysis (FPCA) paradigm to estimate the parameters $\beta(s,t)$ and $\gamma(r,s,t)$ of Model~\eqref{eq:sfofqr}. Moreover, \cite{Jin2023} considered the Model~\eqref{eq:sfofqr} in the framework of reproducing kernel Hibert space. They used the functional ANOVA decomposition to express the three-dimensional model parameter and used a penalized least squares approach to estimate it. Using the Karhunen-Lo\`{e}ve expansions, \cite{LuoQi} extended Model~\eqref{eq:sfofqr} to multiple FoFRM that also considers interactions between multiple functional predictors as follows:
\begin{equation}\label{eq:mfofqr}
\Y_i(t) = \beta_0(t) + \sum_{p=1}^P \int_0^1 \X_{ip}(s) \beta_p(s,t) ds + \sum_{p=1}^P \sum_{p^{\prime}=1}^P \int_0^1 \int_0^1 \X_{ip}(r) \X_{ip^{\prime}}(s) \gamma_{p p^{\prime}}(r,s,t) dr ds + \epsilon_i(t).
\end{equation}
Recently, \cite{BS22} proposed a partial least squares regression to estimate the two- and three-dimensional parameters of Model~\eqref{eq:mfofqr}. The identifiability of Models~\eqref{eq:fof} and~\eqref{eq:sfofqr} is discussed in the supplement.

Compared with the standard FoFRM in~\eqref{eq:fof}, Model~\eqref{eq:mfofqr} is more flexible. The numerical analyses performed by the studies above have proven that the FoFRM with quadratic and interaction effects (FoFRM-QI) produces better prediction results than the standard FoFRM. However, to our knowledge, all the available methods used to estimate two- and three-dimensional parameters of Model~\eqref{eq:mfofqr} are based on non-robust estimation strategies. Their finite-sample performance can be significantly affected in the presence of outliers. While Model~\eqref{eq:mfofqr} has a more flexible form than the FoFRM, the quadratic and interaction effects may be erroneous in the presence of outliers. Outliers in the quadratic and interaction effect terms may affect the model structure and parameter estimation more severely than their main effect. In such cases, the non-robust estimation strategies may produce biased estimates for the model parameters, leading to poor model fitting and predictions.

Inspired by \cite{kalogridis2019} and \cite{Harjit}, we propose a robust FoFRM-QI (RFoFRM-QI) to effectively estimate the coefficient surfaces and hypersurfaces of Model~\eqref{eq:mfofqr} in the presence of outliers. In the proposed method, we first use the robust FPCA (RFPCA) of \cite{Bali2011} to robustly project the infinite-dimensional FoFRM-QI onto a finite-dimensional space of RFPCA bases. The RFPCA of \cite{Bali2011} uses the robust projection pursuit approach of \cite{croux96} combined with a robust scale estimator to produce FPCA estimates. Then, we consider the $\tau$-estimator of \cite{Ben2006} to robustly estimate model parameters constructed via RFPCA basis expansions of functional variables, which are then used to approximate the coefficient surfaces and hypersurfaces. 

Using RFPCA and $\tau$-estimator in the proposed approach results in an estimation strategy robust to outliers in the response variable and the predictor space. The finite-sample performance of the proposed method depends on the number of components extracted from the RFPCA. Similar to \cite{Harjit}, a robust model selection procedure based on the Bayesian information criterion (BIC) is used to determine the optimum number of principal components. 

When several functional predictors are considered in Model~\eqref{eq:mfofqr}, there may be too many quadratic and interaction effect terms to take into account, and not all of them may have a significant effect on the model. In this paper, a forward procedure combined with the robust BIC (RBIC) is considered to determine the model's relevant main, quadratic, and interaction effects (refer to the supplement). Taking advantage of the works \cite{kalogridis2019} and \cite{Harjit}, the asymptotic consistency of the proposed estimators is established under some regulatory conditions (refer to the supplement). In addition, the influence functions of the proposed estimators are derived by combining the influence functions of the RFPCA and $\tau$-estimator. Moreover, the asymptotic distribution of the proposed estimators is derived using the Bahadur expansion.

The remainder of this paper is organized as follows. The RFoFRM-QI and estimation strategy are introduced in Section~\ref{sec:methodology}. The RBIC procedure is discussed in Section~\ref{sec:RBIC}. The superior finite-sample accuracy of the proposed method is demonstrated via a series of Monte-Carlo simulation studies, and the results are presented in Section~\ref{sec:MC}. Finally, Section~\ref{sec:conc} gives some concluding remarks. In a supplement, the asymptotic properties of the proposed method are presented, the variable selection procedure considered in this study is summarized, and the empirical performance of the proposed method is further investigated using a U.S. COVID-19 dataset.

\section{Methodology} \label{sec:methodology}

\subsection{Model, notations, and nomenclature}

Let $\mathcal{H}$ denote a separable Hilbert space defined in $( \Omega, \mathcal{A}, \mathcal{F} )$ with norm $\Vert \alpha \Vert^2 = \langle \alpha, \alpha \rangle$ generated by the inner product $\langle \cdot, \cdot \rangle$. Throughout this study, we consider $\mathcal{L}_2 ( [0,1] ) $ Hilbert space of square-integrable and real-valued functions defined on $[0,1]$, $f: [0,1] \rightarrow \mathbb{R}$ satisfying $\int_0^1 f^2(t) dt < \infty$. In addition, for $\mathcal{L}_2 ( [0,1] ) $, we take the usual inner product, i.e., $\langle f,g \rangle = \int_0^1 f(t) g(t) dt$, $\forall f,g \in \mathcal{L}_2 ( [0,1] )$.

Consider a set of i.i.d. random samples $\lbrace \Y_i(t), \bm{\X}_i(s)$: $i = 1, 2, \ldots n \rbrace$ drawn from a population $\left\lbrace  \Y, \bm{\X} \right\rbrace$ with $\bm{\X} = [ \X_1, \ldots, \X_P ]^\top$. We postulate that $\Y$ and $\X_p$ ($p = 1, \ldots, P$) have second-order moments, i.e., $\text{E}( \Vert \Y \Vert^2 ) = \text{E} ( \Vert \X_p \Vert^2 ) < \infty$. Without loss of generality, we further postulate that the i.i.d. random samples are mean-zero stochastic processes, so that $\text{E}[ \Y(t) ] = \text{E} [ \X_p(s) ] = 0$. Consequently, we consider the following reduced form of FoFRM-QI given in~\eqref{eq:mfofqr}:
\begin{equation}\label{eq:rmfofqr}
\Y_i(t) = \sum_{p=1}^P \int_0^1 \X_{ip}(s) \beta_p(s,t) ds + \sum_{p=1}^P \sum_{p^{\prime}=1}^P \int_0^1 \int_0^1 \X_{ip}(r) \X_{ip^{\prime}}(s) \gamma_{p p^{\prime}}(r,s,t) dr ds + \epsilon_i(t),
\end{equation}
where $\epsilon_i(t) \in \mathcal{L}_2 ( [0,1] )$ is assumed to be independent of $\X_{ip}(s)$ for $p = 1, \ldots, P$ and $\text{E} [\epsilon_i(t) ] =0$, $\forall t \in [0,1]$. 

The elements of functional response and functional predictors in Model~\eqref{eq:rmfofqr} belong to infinite-dimensional $\mathcal{L}_2$ Hilbert space, and the direct estimation of this model is an ill-posed problem. Thus, we first project all the functional objects in~\eqref{eq:rmfofqr} onto basis expansions before fitting the FoFRM-QI. Denote by $\mathcal{C}_{\Y} (t_1, t_2 ) = \text{Cov} [ \Y(t_1), \Y(t_2) ]$ and $\mathcal{C}_{\X_p} (s_1, s_2 ) = \text{Cov} [ \X_p(s_1), \X_p(s_2) ]$ respectively the covariance functions of $\Y(t)$ and $\X_p(s)$ satisfying $\int_0^1 \int_0^1 \mathcal{C}_{\Y}^2 (t_1, t_2 ) dt_1 dt_2 < \infty$ and $\int_0^1 \int_0^1 \mathcal{C}_{\X_p} (s_1, s_2 ) ds_1 ds_2 < \infty$. By Mercer's Theorem, we have the following representations:
\begin{align*}
\mathcal{C}_{\Y} &= \sum_{k \geq 1} w_k \phi_k(t_1) \phi_k(t_2), \qquad \forall t_1, t_2 \in [0,1], \\
\mathcal{C}_{\X_p} &= \sum_{l \geq 1} \kappa_{pl} \psi_{pl}(s_1) \psi_{pl}(s_2), \qquad \forall s_1, s_2 \in [0,1],
\end{align*}
where $\left\lbrace \phi_k(t): k = 1, 2, \ldots \right\rbrace$ and $\left\lbrace \psi_{pl}(s): l = 1, 2, \ldots \right\rbrace$ are orthonormal eigenfunctions in $\mathcal{L}_2 ( [0,1] )$ corresponding to the non-negative eigenvalues $\left\lbrace w_k: k = 1, 2, \ldots \right\rbrace$ and $\left\lbrace \kappa_{pl}: l = 1, 2, \ldots \right\rbrace$ with $w_k \geq w_{k+1}$ and $\kappa_{pl} \geq \kappa_{p(l+1)}$, respectively. We assume that all eigenvalues with respect to $\mathcal{C}_{\Y}$ and $\mathcal{C}_{\X_p}$ are distinct to ensure the uniqueness of orthonormal bases of eigenfunctions. The basis functions are ordered in decreasing order of the corresponding eigenvalues $\left\lbrace w_1 \geq w_2 \geq \ldots \right\rbrace$ and $\left\lbrace \kappa_{p1} \geq \kappa_{p2} \ldots \right\rbrace$. In practice, most of the variability in functional variables can be captured via a finite number of the first few eigenfunctions. Thus, we consider projecting the sample elements of $\Y(t)$ and $\X_p(s)$ onto basis expansions with pre-determined truncation constants $K_{\Y}$ and $K_{\X_p}$, respectively. By Karhunen-Lo\`{e}ve expansion, the elements of functional response and functional predictors can be approximated as follows:
\begin{align}
\Y_i(t) &\approx \sum_{k=1}^{K_{\Y}} \xi_{ik} \phi_k(t) = \bm{\xi}_i^\top \bm{\phi}(t),  \label{eq:by} \\
\X_{ip}(s) &\approx \sum_{l=1}^{K_{\X_p}} \zeta_{ipl} \psi_{pl}(s) = \bm{\zeta}_{ip}^\top \bm{\psi}_p(s), \label{eq:bx}
\end{align}
where the random variables $\xi_{ik} = \int_0^1 \Y_i(t) \phi_k(t) dt$ and $\zeta_{ipl} = \int_0^1 \X_{ip}(s) \psi_{pl}(s) ds$ are the projections of $\Y_i(t)$ and $\X_{ip}(s)$ onto their corresponding orthonormal bases, respectively.

The two- and three-dimensional coefficient surfaces and hypersurfaces also admit a similar representation:
\begin{align}
\beta_p(s,t) &\approx \sum_{l=1}^{K_{\X_p}} \sum_{k=1}^{K_{\Y}} b_{plk} \psi_{pl}(s) \phi_k(t) = \bm{\psi}_p^\top (s) \bm{B}_p \bm{\phi}(t), \qquad \forall s, t \in [0,1], \label{eq:bb} \\
\gamma_{p p^{\prime}}(r,s,t) &\approx \sum_{l=1}^{K_{\X_p}} \sum_{m=1}^{K_{\X_p}} \sum_{k=1}^{K_{\Y}} \gamma_{p p^{\prime} l m k} \psi_{pl}(r) \psi_{p^{\prime} m} (s) \phi_k(t), \nonumber\\
& = [ \bm{\psi}_{p^{\prime}}(s) \otimes \bm{\psi}_p(r) ]^\top \bm{\Gamma}_{p p^{\prime} (3)}^\top \bm{\phi}(t), \qquad \forall  r, s, t \in [0,1] \label{eq:bg},
\end{align}
where $b_{plk} = \int_0^1 \int_0^1 \beta_p(s,t) \psi_{pl}(s) \phi_k(t) ds dt$, $\gamma_{p p^{\prime} l m k} = \int_0^1 \int_0^1 \int_0^1  \gamma_{p p^{\prime}} (r,s,t) \psi_{pl}(r) \psi_{p^{\prime} m} (s) \\ \phi_k(t) dr ds dt$, and $\bm{\Gamma}_{p p^{\prime} (3)}$ is a $K_{\X_p}^2 \times K_{\Y}$ dimensional matrix obtained by matricizing a three-dimensional $K_{\X_p} \times K_{\X_p} \times K_{\Y}$ tensor $\Gamma = ( \gamma_{p p^{\prime} l m k} )_{lmk}$ with respect to the $3\textsuperscript{rd}$ array \citep[see, e.g.,][]{Lathuawer, Matsui2020}.

Let the error function $\epsilon_i(t)$ admit the basis expansion with the same basis functions in $\Y_i(t)$:
\begin{equation}
\epsilon_i(t) = \sum_{k=1}^{K_{\Y}} e_{ik} \phi_k(t) = \bm{e}_i^\top \bm{\phi}(t), \qquad \forall t \in [0,1] \label{eq:be},
\end{equation}
where the random variable $e_{ik} = \int_0^1 \epsilon_i(t) \phi_k(t)$ is typically assumed to follow a normal distribution. Then, substituting~\eqref{eq:by}-\eqref{eq:be} in~\eqref{eq:rmfofqr}, we have
\begin{align*}
\bm{\xi}_i^\top \bm{\phi}(t) &= \sum_{p=1}^P \bm{\zeta}_{ip}^\top \bm{\psi}_p(s) \bm{\psi}_p^\top(s) \bm{B}_p \bm{\phi}(t) \\
& + \sum_{p=1}^P \sum_{p^{\prime}=1}^P \bm{\zeta}_{ip}^\top \bm{\psi}_p(r) \bm{\zeta}_{ip^{\prime}}^\top \bm{\psi}_{p^{\prime}}(s) [  \bm{\psi}_{p^{\prime}}(s) \otimes \bm{\psi}_p(r) ]^\top \bm{\Gamma}_{p p^{\prime} (3)}^\top \bm{\phi}(t) + \bm{e}_i^\top \bm{\phi}(t).
\end{align*}
By orthonormalities of $\bm{\phi}(t)$, $\bm{\psi}_p(r)$, and $\bm{\psi}_{p^{\prime}}(s)$, i.e., $\int_0^1 \bm{\phi}(t) \bm{\phi}^\top(t) = 1$ (similarly for $\bm{\psi}_p(r)$ and $\bm{\psi}_{p^{\prime}}(s)$), the FoFRM-QI is reduced to the following form:
\begin{equation}\label{eq:redf}
\bm{\xi}_i^\top = \sum_{p=1}^P \bm{\zeta}_{ip}^\top \bm{B}_p + \sum_{p=1}^P \sum_{p^{\prime}=1}^P ( \bm{\zeta}_{ip} \otimes \bm{\zeta}_{ip^{\prime}} )^\top \bm{\Gamma}_{p p^{\prime} (3)}^\top + \bm{e}_i^\top.
\end{equation}

We obtain robust estimates of two-dimensional coefficient surfaces $\beta(s,t)$ and three-dimensional hypersurfaces $\gamma_{p p^{\prime}}(s,r,t)$ in~\eqref{eq:mfofqr} by robustly estimating all coefficients in~\eqref{eq:redf}. In doing so, we consider the RFPCA of \cite{Bali2011} and the $\tau$-estimator of \cite{Ben2006}.

\subsection{The RFPCA estimates of basis expansion coefficients}

RFPCA works similarly to the classical FPCA but uses a robust scale functional in calculations instead of variance. We begin by denoting the set of all univariate distributions by $\mathcal{G}$. The location invariant and scale invariant scale functional, denoted by $\sigma_R$, is defined as $\sigma_R: \mathcal{G} \rightarrow [ 0, + \infty)$. This study considers the M-scale functional $\sigma_M(G)$ where $G$ denotes the distribution of a given random variable $Z$. The M-scale functional can be obtained as a solution of a loss-function $\rho: \mathbb{R} \rightarrow \mathbb{R}$ as follows:
\begin{equation*}
\text{E} \left[ \rho \left( \frac{Z - \mu}{\sigma_R(G)} \right) \right] = \delta,
\end{equation*}
where $\mu$ is a location parameter. We assume that the loss-function $\rho$ satisfies the following properties \citep{Ben2006}:
\begin{asparaenum}
\item[C1.] $\rho$ is a continuous function.
\item[C2.] $\rho(0) = 0$.
\item[C3.] $0 \leq u \leq u^*$ implies $\rho(u) \leq \rho(u^*)$.
\item[C4.] $0 < \text{A} = \sup_u \rho(u) < \infty$.
\item[C5.] If $\rho(v) < \text{A}$ and $0 \leq v < u$, then $\rho(v) < \rho(u)$.
\end{asparaenum}
For a sample $\bm{z} = \left\lbrace z_1, \ldots, z_n \right\rbrace$ and its location estimate $\widehat{\mu}_n$, the sample M-estimate of scale $\widehat{\sigma}_n$ is obtained as solution to the M-estimating equation
\begin{equation}\label{eq:mscale}
\frac{1}{n} \sum_{i=1}^n \rho \left( \frac{z_i - \widehat{\mu}}{\widehat{\sigma}_n} \right) = \delta.
\end{equation}
The M-estimate of scale $\widehat{\sigma}_n$ converges to its true value, denoted by $\sigma$ (which is defined by $\delta = \text{E}_G [ \rho (z / \sigma ) ]$), when $z$ is an i.i.d. random sample from $G$. Note that if $\delta = \text{E}_G [ \rho \left(z \right) ]$, then the M-scale is calibrated so that $\sigma = 1$. Similar to \cite{Bali2011}, we consider the loss-function proposed by \cite{Beaton1974} as follows:
\begin{equation}\label{eq:loss1}
        \rho_{0,c}(u) =
        \left\{ \begin{array}{ll}
            \frac{u^2}{2} \left( 1 - \frac{u^2}{c^2} + \frac{u^4}{3c^4} \right)  & \text{if}~ \vert u \vert \leq c, \\
            \frac{c^2}{6} & \text{if}~ \vert u \vert > c,
        \end{array} \right.
\end{equation}
where $c$ is the tuning parameter that controls the robustness and efficiency of the estimator. The popular choices of $c$ and $\delta$, used in our numerical analyses for the RFPCA estimates, are $c = 1.56$ and $\delta = 1/2$. With these choices, the M-scale functional is Fisher-consistent at the normal distribution and has a 50\% breakdown point \citep{Bali2011}.

Based on the definitions given above, we summarize how the RFPCA estimates of the coefficients in~\eqref{eq:redf} are obtained. For simplicity, we consider only the RFPCA estimates of the functional response $\Y(t)$. The estimates for the functional predictors $\X_p(s)$ follow the same lines as for $\Y(t)$. Denote by $\mathcal{F}[\alpha]$ the distribution of $\langle \alpha, \Y \rangle$ when the distribution of $\Y$ is $\mathcal{F}$, i.e., $\Y \sim \mathcal{F} $. For a given M-scale functional $\sigma_M(\mathcal{F})$, the orthonormal bases of eigenfunctions are defined as follows:
\[ \begin{cases}
\phi_k(\mathcal{F}) = \underset{\begin{subarray}{c} 
      \Vert \alpha \Vert^2 = 1
\end{subarray}}{\argmax}~ \sigma_M(\mathcal{F}[\alpha]), & k = 1, \\
\phi_k(\mathcal{F}) = \underset{\begin{subarray}{c} 
      \Vert \alpha \Vert^2 = 1, \alpha \in \mathcal{B}_k
\end{subarray}}{\argmax}~ \sigma_M(\mathcal{F}[\alpha]), & k \geq 2,
   \end{cases}
\]
where $\mathcal{B}_k = \left\lbrace \alpha \in \mathcal{L}_2 ( [0,1] ): \langle \alpha, \phi_k ( \mathcal{F} ) \rangle = 0, ~ 1 \leq k \leq \ K_{\Y} - 1 \right\rbrace$. In addition, the $k^\textsuperscript{th}$ largest eigenvalue is defined by:
\begin{equation*}
w_k(\mathcal{F}) = \sigma_M^2(\mathcal{F}[\phi_k]) = \underset{\begin{subarray}{c} 
      \Vert \alpha \Vert^2 = 1, \alpha \in \mathcal{B}_k
\end{subarray}}{\max} \sigma_M^2(\mathcal{F}[\alpha]).
\end{equation*}

Let $\sigma_M(\mathcal{F}_n[\alpha])$ denote the functional for $\sigma_M$ obtained using the empirical distribution of $\left\lbrace \langle \alpha, \Y_1 \rangle, \ldots, \langle \alpha, \Y_n \rangle \right\rbrace $. Let also $s^2_n: \mathcal{L}_2 ( [0,1] ) \rightarrow \mathbb{R}$ denote the function of empirical M-scale functional such that $s^2(\alpha) = \sigma_M^2(\mathcal{F}[\alpha])$. Then, the RFPCA estimates of the orthonormal bases of eigenfunctions for $\Y(t)$ are given by
\[ \begin{cases}
\widehat{\phi}_k(t) = \underset{\begin{subarray}{c} 
      \Vert \alpha \Vert^2 = 1
\end{subarray}}{\argmax}~ s_n(\alpha), & k = 1, \\
\widehat{\phi}_k(t) = \underset{\begin{subarray}{c} 
      \alpha \in \widehat{\mathcal{B}}_k
\end{subarray}}{\argmax}~ s_n(\alpha), & k \geq 2,
   \end{cases}
\]
where $\widehat{\mathcal{B}}_k = \left\lbrace \alpha \in \mathcal{L}_2 ( [0,1] ): \Vert \alpha \Vert = 1, \langle \alpha, \widehat{\phi}_k \rangle = 0, ~ \forall~ 1 \leq k \leq \ K_{\Y} - 1 \right\rbrace$. Accordingly, the corresponding eigenvalues are obtained as follows:
\begin{equation*}
\widehat{w}_k = s^2_n (\widehat{\phi}_k), \quad k \geq 1.
\end{equation*}

Let $[ \widehat{\xi}_{ik}, \widehat{\phi}_k(t) ]$ and $[ \widehat{\zeta}_{ipl}, \widehat{\psi}_{pl}(s)]$ denote the RFPCA estimates of $[ \xi_{ik}, \phi_k(t) ]$ and $[ \zeta_{ipl}, \psi_{pl}(s) ]$, respectively. Then, we represent the multivariate regression problem in~\eqref{eq:redf} as follows:
\begin{equation}\label{eq:rform}
\bm{\Xi} = \bm{\Pi} ~ \bm{\Theta} + \bm{e},
\end{equation}
where $\bm{\Xi} = [ \widehat{\bm{\xi}}_1^\top, \ldots, \widehat{\bm{\xi}}_n^\top ]^\top$ with $\widehat{\bm{\xi}}_i = [ \widehat{\xi}_{i1}, \ldots, \widehat{\xi}_{iK_{\Y}} ]^\top$ is a $K_{\Y}$-dimensional matrix, $\bm{\Pi}$ is $( \sum_{p=1}^P K_{\X_p} + \sum_{p=1}^p \sum_{p^{\prime}=1}^P K_{\X_p} \times K_{\X_{p^{\prime}}} )$-dimensional matrix consisting of both main and interaction effects terms such that $\bm{\Pi} = [ ( \widehat{\bm{\zeta}}_1^\top \ldots, \widehat{\bm{\zeta}}_n^\top), [ ( \widehat{\bm{\zeta}}_1^{\otimes} )^\top, \ldots, ( \widehat{\bm{\zeta}}_n^{\otimes} )^\top] ]^\top $ where $\widehat{\bm{\zeta}}_i = [ \widehat{\bm{\zeta}}_{i1}^\top, \ldots, \widehat{\bm{\zeta}}_{iP}^\top ]^\top$, $\widehat{\bm{\zeta}}_{ip} = [ \widehat{\zeta}_{ip1}, \ldots, \widehat{\zeta}_{ipK_{\X_p}} ]^\top$, and $\widehat{\bm{\zeta}}_i^{\otimes} = \widehat{\bm{\zeta}}_{ip} \otimes \widehat{\bm{\zeta}}_{ip^{\prime}}$, $\bm{\Theta} = [ \bm{B} ~ ~ \bm{\Gamma}^\top ]^\top$ with $\bm{B} = [ \bm{B}_1, \ldots, \bm{B}_p ]^\top$ and $\bm{\Gamma} = \left\lbrace \bm{\Gamma}_{pp^{\prime}}: p, p^{\prime} = 1, \ldots, P \right\rbrace$ is $( \sum_{p=1}^P K_{\X_p} + \sum_{p=1}^p \sum_{p^{\prime}=1}^P K_{\X_p} \times K_{\X_{p^{\prime}}} ) \times K_{\Y}$ dimensional parameter matrix, and $\bm{e}$ = $[ \widehat{\bm{e}}_1^{\top}, \ldots, \widehat{\bm{e}}_n^{\top} ]^{\top}$ with $\bm{e}_i = [ \widehat{e}_{i1}, \ldots, \widehat{e}_{iK_{\Y}} ]^{\top}$. We typically assume that $\bm{e}$ follows a normal distribution with mean-zero and variance-covariance matrix $\bm{\Sigma}$, i.e., $\bm{e} \sim \mathcal{N} ( \bm{0}, \bm{\Sigma} )$.

\subsection{$\tau$ estimates of the multivariate regression parameters}

With basis expansions of functional objects, the problem of estimating infinite-dimensional model parameters $\beta_p(s,t)$ and $\gamma_{p p^{\prime}}$ is reduced to the problem of estimating the parameters of $\bm{\Theta}$ in the space of expansion coefficients $\bm{\Xi}$ and $\bm{\Pi}$ in the multivariate regression model of~\eqref{eq:rform}. Several methods have been proposed to robustly estimate the parameters of a multivariate linear regression model, such as the M-estimator based on a convex $\rho$-function \citep{Koenker1990}, S-estimator \citep{Bilodeau2000}, the minimum covariance determinant estimator \citep{Rousseeuw2004}, and the multivariate least trimmed squares estimator \citep{Agullo2008}. While these estimators have a high breakdown point, they are not highly efficient when $\bm{e}$ is normal \citep{Ben2006, kudraszow2011}. This paper considers the robust estimation procedure based on the $\tau$-scale proposed by \cite{Ben2006} to obtain a robust and efficient estimate for $\bm{\Theta}$. Compared with other available robust estimators, the $\tau$-estimator has a high breakdown point and high normal efficiency. A high breakdown point indicates that the $\tau$-estimator has a high degree of robustness to outliers. In contrast, high normal efficiency indicates that the variance of the $\tau$-estimator is not much larger than those of the least-squares estimator under the normally distributed regression model. In addition, the numerical analyses performed by \cite{Dutta2017} have shown that the $\tau$-estimator yields the best overall performance among the existing multivariate robust approaches.

For a regression model, \cite{Hosser1992} showed that the parameter estimates based on the M-scale could not combine the high breakdown point and high-efficiency properties. Thus, we consider the $\tau$-estimates of the scale proposed by \cite{yohai88}, which can provide regression estimates with a high breakdown point and asymptotic efficiency under normality. Similar to~\eqref{eq:loss1}, let us consider two loss-functions $\rho_{0,c}$, $\rho_{1,c}$ satisfying C1-C5 and let $\delta_i = \text{E}_G [ \rho_{i,c} ( z ) ]$ for $i = 0, 1$. Let $\widehat{\sigma}_n$ denote the M-estimate of scale~\eqref{eq:mscale} obtained via $ \rho_{0,c}$ and $\delta_0$. Then, the $\tau$-estimate of scale for the sample $z$ is defined as follows:
\[
\tau^2(z) = \frac{\widehat{\sigma}_n^2}{n} \sum_{i=1}^n \rho_{1,c} \left( \frac{z_i}{\widehat{\sigma}_n}\right).
\]
Let $\varphi_i(z) = \partial \rho_{i,c}(z) / \partial z$. We assume that $\rho_{1,c}$ is continuously differentiable and satisfies the following property:
\begin{asparaenum}
\item[C6.] $2 \rho_{1,c}(u) - \varphi_1(u) u$ for $u > 0$,
\end{asparaenum}
which guarantees the Fisher-consistency of the $\tau$-estimates of regression \citep{Ben2006}. For any suitable matrices $\bm{u}$ and $\bm{v}$, let us denote by $d_i ( \bm{u}, \bm{v} )  = ( \bm{u}^\top \bm{v}^{-1} \bm{u} )$ the Mahalanobis norm of $\bm{u}$ with respect to $\bm{v}$. Then, similar to \citep{Ben2006}, we define the $\tau$-estimates of Model~\eqref{eq:rform}, denoted by $\widehat{\bm{\Theta}}_{\tau} = [ \widehat{\bm{B}}_{\tau} ~~ \widehat{\bm{\Gamma}}_{\tau} ]$ and $\widehat{\bm{\Sigma}}_{\tau}$, by
\begin{equation*}
\left[\widehat{\bm{\Theta}}_{\tau}, \widehat{\bm{\Sigma}}_{\tau} \right] = \underset{\begin{subarray}{c} 
      \bm{\Theta}, \bm{\Sigma}
\end{subarray}}{\argmin} ~~ \text{det} ( \bm{\Sigma} ), 
\end{equation*}
subject to $\tau^2 [ d_1 ( \bm{\Theta}, \bm{\Sigma} ), \ldots, d_n ( \bm{\Theta}, \bm{\Sigma} ) ]$. If we denote the true parameter matrices by $\bm{\Theta}_0 = [ \bm{B}_0 ~~ \bm{\Gamma}_0 ]$ and $\bm{\Sigma}_0$, the $\tau$ estimate $\widehat{\bm{\Theta}}_{\tau}$ converges to $\bm{\Theta}_0$ when the true distribution of errors is elliptical such as multivariate normal \citep[see, e.g.,][]{Ben2006}.

Let $d_i^*$, $C_n = C_n ( \bm{\Theta}, \bm{\Sigma} )$, $D_n = D_n ( \bm{\Theta}, \bm{\Sigma} )$, $\varphi_n^* = \varphi^*_{n, \bm{\Theta}, \bm{\Sigma}}$, and $w_n^*$ be defined as follows:
\begin{align*}
d_i^*( \bm{\Theta}, \bm{\Sigma} ) &= \frac{d_i ( \bm{\Theta}, \bm{\Sigma} )}{s[d_1( \bm{\Theta}, \bm{\Sigma} ), \ldots, d_n( \bm{\Theta}, \bm{\Sigma} )]} \\
C_n &= \frac{1}{n} \sum_{i=1}^n [ 2 \rho_{1,c} [ d_i^*( \bm{\Theta}, \bm{\Sigma} ) ] - \varphi_1 [ d_i^*( \bm{\Theta}, \bm{\Sigma} ) ] d_i^*( \bm{\Theta}, \bm{\Sigma} ) ] \\
D_n &= \frac{1}{n} \sum_{i=1}^n [ \varphi_0 [ d_i^*( \bm{\Theta}, \bm{\Sigma} ) ] d_i^*( \bm{\Theta}, \bm{\Sigma} ) ] \\
\varphi^*_n ( z ) &= C_n \varphi_0 ( z ) + D_n \varphi_1 ( z ) \\
w^*_n &= \frac{\varphi^*_n ( z )}{z},
\end{align*}
where $s[d_1( \bm{\Theta}, \bm{\Sigma} ), \ldots, d_n( \bm{\Theta}, \bm{\Sigma} )]$ denotes the M-estimates of scale. Then, we obtain the $\tau$-estimates $[ \widehat{\bm{\Theta}}_{\tau}, \widehat{\bm{\Gamma}}_{\tau} ]$ by solving the following estimating equations:
\begin{align}
\sum_{i=1}^n w_n^* [ d_i^* ( \widehat{\bm{\Theta}}_{\tau}, \widehat{\bm{\Sigma}}_{\tau} ) ] \widehat{e}_i ( \widehat{\bm{\Theta}}_{\tau} ) \bm{\Pi}_i &= 0 \label{eq:taub} \\
\widehat{\bm{\Sigma}}_{\tau} &= \frac{K_{\Y} \sum_{i=1}^n w^*_n [ d_i^* ( \widehat{\bm{\Theta}}_{\tau}, \widehat{\bm{\Sigma}}_{\tau} ) ] \widehat{e}_i ( \widehat{\bm{\Theta}}_{\tau} ) \widehat{e}_i^\top ( \widehat{\bm{\Theta}}_{\tau} )}{\widetilde{s}^2 \sum_{i=1}^n \varphi^*_n [ d_i^* ( \widehat{\bm{\Theta}}_{\tau}, \widehat{\bm{\Sigma}}_{\tau} ) ] d_i^* ( \widehat{\bm{\Theta}}_{\tau}, \widehat{\bm{\Sigma}}_{\tau} )}, \label{eq:taus}
\end{align}
where $\widehat{\bm{e}}_i ( \widehat{\bm{\Theta}}_{\tau} ) = \bm{\xi}_i^\top - \bm{\Pi} \widehat{\bm{\Theta}}_{\tau}$ and $\widetilde{s}^2 = s [ d_1 ( \widehat{\bm{\Theta}}_{\tau}, \widehat{\bm{\Sigma}}_{\tau} ), \ldots, d_n ( \widehat{\bm{\Theta}}_{\tau}, \widehat{\bm{\Sigma}}_{\tau} ) ]$.

From~\eqref{eq:taub} and~\eqref{eq:taus}, the estimates $[ \widehat{\bm{\Theta}}_{\tau}, \widehat{\bm{\Sigma}}_{\tau} ]$ are computed based on the weights $w_n^* [ d_i^* ( \widehat{\bm{\Theta}}_{\tau}, \widehat{\bm{\Sigma}}_{\tau} ) ]$. However, these weights depend on the estimates $[ \widehat{\bm{\Theta}}_{\tau}, \widehat{\bm{\Sigma}}_{\tau} ]$. Thus, the estimates are not obtained directly but computed via an iterative algorithm. Let $ \widehat{\bm{\Theta}}^{(0)}$ and $\widehat{\bm{\Sigma}}^{(0)}$ denote the initial estimates at the beginning of the algorithm. The  M-estimate of scale $s^{(0)} = s [ d_1 ( \widehat{\bm{\Theta}}^{(0)}, \widehat{\bm{\Sigma}}^{(0)} ) ] $ is computed using the initial estimates to obtain the weights $w^*_n [ d_i ( \widehat{\bm{\Theta}}^{(0)}, \widehat{\bm{\Sigma}}^{(0)} ) / s^{(0)} ]$. These weights are then used to compute $ \widehat{\bm{\Theta}}^{(1)}$ by the weighted least squares. Based on $ \widehat{\bm{\Theta}}^{(1)}$, $\widehat{\bm{\Sigma}}^{(1)}$ is computed as $\widehat{\bm{\Sigma}}^{(1)} = [ ( \widehat{\tau}^{(1)} )^2 / \delta_1 ] \widehat{\bm{\Sigma}}^{*(1)}$ where
\begin{equation*}
\widehat{\bm{\Sigma}}^{*(1)} = \frac{K_{\Y} \sum_{i=1}^n w^*_n \left[ \frac{d_i ( \widehat{\bm{\Theta}}^{(1)}, \widehat{\bm{\Sigma}}^{(0)})}{s^{(1)}} \right] e_i (\widehat{\bm{\Theta}}^{(1)}) e_i^\top (\widehat{\bm{\Theta}}^{(1)}) }{( s^{(1)} )^2 \sum_{i=1}^n \varphi^*_n \left[\frac{d_i ( \widehat{\bm{\Theta}}^{(1)}, \widehat{\bm{\Sigma}}^{(0)})}{s^{(1)}} \right] \frac{d_i ( \widehat{\bm{\Theta}}^{(1)}, \widehat{\bm{\Sigma}}^{(0)})}{s^{(1)}}},
\end{equation*}
where $s^{(1)} = s [ d_1 ( \widehat{\bm{\Theta}}^{(1)}, \widehat{\bm{\Sigma}}^{(0)}), \ldots, d_n ( \widehat{\bm{\Theta}}^{(1)}, \widehat{\bm{\Sigma}}^{(0)}) ]$ and $\widehat{\tau}^{(1)} = \tau [ d_1 ( \widehat{\bm{\Theta}}^{(1)}, \widehat{\bm{\Sigma}}^{*(1)} ), \ldots,\\ d_n ( \widehat{\bm{\Theta}}^{(1)}, \widehat{\bm{\Sigma}}^{*(1)} ) ]$. Denote by $ \widehat{\bm{\Theta}}^{(h)}$ and $\widehat{\bm{\Sigma}}^{(h)}$ the estimates computed at $h^\textsuperscript{th}$ iteration using $ \widehat{\bm{\Theta}}^{(h-1)}$ and $\widehat{\bm{\Sigma}}^{(h-1)}$ as the initial estimates. The algorithm stops at iteration $H$ if convergence is achieved. Convergence is reached if the relative absolute difference between all the elements of $ \widehat{\bm{\Theta}}^{(H)}$ and $ \widehat{\bm{\Theta}}^{(H-1)}$ are smaller than a predetermined constant $\varepsilon$. The initial estimates at the beginning of the algorithm are computed by a subsampling approach \citep[see, e.g.,][for more details]{Ben2006}.

Finally, let $\widehat{\bm{\phi}}(t) = [ \widehat{\phi}_1(t), \ldots, \widehat{\phi}_{K_{\Y}}(t) ]^\top$, $\widehat{\bm{\psi}}_p(s) = [ \widehat{\psi}_{p1}(s), \ldots, \widehat{\psi}_{p K_{\X_p}}(s) ]^\top$, and $\widehat{\bm{\psi}}_{p^{\prime}}(s) = [ \widehat{\psi}_{p^{\prime}1}(s),$ $\ldots, \widehat{\psi}_{p^{\prime} K_{\X_p}}(s) ]^\top$ denote the RFPCA estimates of $\bm{\phi}(t)$, $\bm{\psi}(s)$, and $\bm{\psi}(r)$, respectively. In addition, let $\widehat{\bm{\Theta}}_{\tau} = [ \widehat{\bm{B}}_{\tau} ~~ \widehat{\bm{\Gamma}}_{\tau} ]$ with $\widehat{\bm{B}}_{\tau} = [ \widehat{\bm{B}}_{\tau 1}, \ldots, \widehat{\bm{B}}_{\tau P} ]^\top$ and $\widehat{\bm{\Gamma}}_{\tau} = \left\lbrace \widehat{\bm{\Gamma}}_{\tau p p^{\prime}}: p, p^{\prime} = 1, \ldots, P \right\rbrace $ denote the $\tau$-estimates of $\bm{\Theta} = [ \bm{B} ~~ \bm{\Gamma} ]$. Then, the robust estimates of two- and three-dimensional coefficient surfaces and hypersurfaces are obtained as follows:
\begin{align}
\widehat{\beta}_{\tau p}(s,t) &= \widehat{\bm{\psi}}_p^\top (s) \widehat{\bm{B}}_{\tau p} \widehat{\bm{\phi}}(t), \label{eq:estb} \\
\widehat{\gamma}_{\tau p p^{\prime}}(r,s,t) &= [ \widehat{\bm{\psi}}_{p^{\prime}}(r) \otimes \widehat{\bm{\psi}}_p(s) ]^\top \widehat{\bm{\Gamma}}_{\tau p p^{\prime} (3)}^\top \widehat{\bm{\phi}}(t). \label{eq:estg}
\end{align}

Note that the FoFRM-QI in~\eqref{eq:rmfofqr} with a large set of functional predictors results in excessive amounts of quadratic and interaction effects, making it difficult to use the proposed method to estimate the model parameters efficiently. To overcome this problem, we consider an extended version of the variable selection procedure introduced by \cite{BS22} (refer to the supplement for more details).

\section{\large{The RBIC for determination of optimum $K_{\Y}$ and $K_{\X_p}$}}\label{sec:RBIC}

The performance of the proposed method depends on the truncation constants $K_{\Y}$ and $K_{\X_p}$ for $p = 1, \ldots, P$. Since the two- and three-dimensional coefficients surfaces and hypersurfaces depend on $K_{\Y}$ and $K_{\X_p}$, we consider jointly determining their optimum values using BIC. The BIC is computed based on the $\log$-likelihood, expressed via the squared error terms. However, the classical BIC is non-robust to outliers and can lead to incorrect determination of optimum truncation constants. Thus, we consider the RBIC presented by \cite{Harjit} to robustly determine the optimum values of $K_{\Y}$ and $K_{\X_p}$.

Let $\bm{Y}_i$ and $\bm{E}_i$ respectively denote the values of $\Y_i(t)$ and $\epsilon_i(t)$ recorded at discrete time points $t = \left\lbrace t_1, \ldots, t_{J_{\Y}} \right\rbrace$, i.e., $\bm{Y}_i = [ \Y_i(t_1), \ldots, \Y_i(t_{J_{\Y}}) ]$ and $\bm{E}_i = [ \epsilon_i(t_1), \ldots, \epsilon_i(t_{J_{\Y}}) ]$. Herein, for simplicity, $\bm{E}_i$ is assumed to be sampled from $\mathcal{N} (0, \sigma^2 \bm{I})$ where $\bm{I}$ denotes the identity matrix. Let us denote the set of all possible models by $\mathcal{J} = \left\lbrace K_{\Y}, K_{\X_p} \vert K_{\Y} = 1, \ldots, K_{\Y_{\max}}, K_{\X_p} = 1, \ldots, K_{\X_{p_{\max}}} \right\rbrace$. Then, it is assumed that there exists a true model associated with $K_{\Y_0}, K_{\X_{p_0}} \in \mathcal{J}$ as follows:
\begin{equation}\label{eq:trm}
\bm{Y}_i = \bm{\Pi}^{(K_{\X_0})}_i ~ \bm{\Theta}^{(K_{\Y_0}, K_{\X_0})} \bm{\phi}^{(K_{\Y_0})} + \bm{E}_i,
\end{equation}
where $\bm{\Pi}^{(K_{\X_0})}_i$, $\bm{\Theta}^{(K_{\Y_0}, K_{\X_0})}_i$, and $\bm{\phi}^{(K_{\Y_0})}_i$ are obtained with truncation constants $K_{\Y_0}$ and $K_{\X_0} = \left\lbrace K_{\X_{1_0}}, \ldots, K_{\X_{P_0}} \right\rbrace$. For computational simplicity, we apply the same truncation constant $K_{\X}$ to all the functional predictors. For each model constructed based on the constants $( K_{\Y}, K_{\X} ) \in \mathcal{J}$, the likelihood function for $i^\textsuperscript{th}$ sample is defined by:
\begin{equation}\small\label{eq:lhood}
f ( \bm{Y}_i \vert \bm{\Delta}^{(K_{\Y}, K_{\X})} ) = \frac{1}{(2 \pi)^{n/2} ( \sigma^{(K_{\Y}, K_{\X})} )^n} \exp \left\lbrace - \frac{[ \bm{Y}_i - \widehat{\bm{Y}}_i^{(K_{\Y}, K_{\X})} ]^\top [ \bm{Y}_i - \widehat{\bm{Y}}_i^{(K_{\Y}, K_{\X})} ]}{2 ( \sigma^{(K_{\Y}, K_{\X})} )^2} \right\rbrace,
\end{equation}
where $\bm{\Delta}^{(K_{\Y}, K_{\X})} = [ \bm{B}^{(K_{\Y}, K_{\X})}, \bm{\Gamma}^{(K_{\Y}, K_{\X})}, \sigma^{(K_{\Y}, K_{\X})} ]$ is the vector of model parameters and $\widehat{\bm{Y}}_i^{(K_{\Y}, K_{\X})}$ is the prediction of $\bm{Y}_i$ computed using~\eqref{eq:trm} with $( K_{\Y}, K_{\X} )$.

Denote by $[\cdot]: \mathbb{R} \rightarrow \mathbb{Z}$ the function that rounds its elements to the nearest integer and defines the set $\mathcal{Z} = \left\lbrace \mathcal{Z} \subset \left\lbrace 1, \ldots, n \right\rbrace, \vert \mathcal{Z} \vert = [\alpha n] \right\rbrace$ for a given $\alpha \in [0,1]$. Then, similar to \cite{Harjit}, we define the trimmed $\log$-likelihood for~\eqref{eq:lhood} by $\ell_{\mathcal{Z}} ( \bm{\Delta}^{(K_{\Y}, K_{\X})}, \mathcal{Z} ) = \sum_{i \in \mathcal{Z}} \ln [ f ( \bm{Y}_i \vert \bm{\Delta}^{(K_{\Y}, K_{\X})} ) ]$. Let $Z^{(K_{\Y}, K_{\X})} = \underset{\rm Z \in \mathcal{Z} }{\rm \argmin}~ \ell_{\mathcal{Z}} ( \bm{\Delta}^{(K_{\Y}, K_{\X})}, Z )$. Then, the RBIC for a given $( K_{\Y}, K_{\X} )$ is defined by:
\begin{equation}\label{eq:rbic}
\text{RBIC} ( K_{\Y}, K_{\X} ) = -2 \ell_{\mathcal{Z}} ( \bm{\Delta}^{(K_{\Y}, K_{\X})}, Z^{(K_{\Y}, K_{\X})} ) + \omega ( K_{\Y}, K_{\X} ) \ln ( [\alpha n] ),
\end{equation}
where $\omega ( K_{\Y}, K_{\X} ) = K_{\Y} \times K_{\X}+1$. In our numerical calculations, we take $\alpha = 0.8$ \citep[i.e., 20\% trimming proportion, which is commonly used in robust statistics, see, e.g.,][]{Wilcox}. In a nutshell, we build $K_{\Y_{\max}} \times K_{\X_{\max}}$ different model (i.e., $K_{\Y} = 1, \ldots, K_{\Y_{\max}}$ and $K_{\X} = 1, \ldots, K_{\X_{\max}}$), where $K_{\Y_{\max}}$ and $K_{\X_{\max}}$ are chosen to ensure that at least 90\% of the variation in the discretely observed data is captured. Then, the optimum $K_{\Y}$ and $K_{\X}$, denoted by $\widetilde{K}_{\Y}$ and $\widetilde{K}_{\X}$ are determined by $( \widetilde{K}_{\Y}, \widetilde{K}_{\X}) = \underset{\rm K_{\Y}, K_{\X} }{\rm \argmin}~~ \text{RBIC} ( K_{\Y}, K_{\X} )$. Under the assumption that $\Y(t)$ and $\X(s)$ have a finite-dimensional Karhunen Lo\`{e}ve decomposition, i.e., $w_k = \kappa_{l} = 0$ for $k > K_{\Y_0}$ and $ l > K_{\X_{0}}$, \cite{Harjit} showed that the estimated truncation constants $( \widetilde{K}_{\Y}, \widetilde{K}_{\X})$ are consistent estimators of $( K_{\Y_0}, K_{\X_{0}} )$ \citep[see e.g., Theorem 4.1 of][for more details]{Harjit}.

\section{Numerical Results} \label{sec:MC}

In this section, the finite-sample performance of the proposed RFoFRM-QI is evaluated via a series of Monte Carlo experiments. Because of limited space, the empirical data analysis results are presented in the supplement. In numerical analyses, the performance of the proposed method is compared with the standard FoFRM-QI (which is estimated via the traditional FPCA and least-squares estimator) and the method proposed by \cite{LuoQi} (``LQ'', hereafter). In addition, the results of the proposed method obtained using only the main effect terms are compared with those of penalized function-on-function regression (``pffr'') method proposed by \cite{ivanescu2015} and the boosting functional regression model (``FDB'') proposed by \cite{FDboost}. An example \Rlogo \ code for all the methods considered in this study can be found at \url{https://github.com/UfukBeyaztas/RFoFRM-QI}. 

The regression model is generated throughout the experiments using several relevant and redundant functional predictor variables with three different sample sizes, $n = [100,~ 250,~ 500]$. In addition, three different contamination levels (CL), $\text{CL} = [0\%,~5\%,~10\%]$, where $0\%$ corresponds to outlier-free data case, are considered to investigate the robustness of the proposed method. The following procedure is repeated 250 times to compare the estimation and predictive performance of the RFoFRM-QI with the FoFRM-QI, LQ, pffr, and FDB. The methods are used to estimate the model parameters with the generated data, and the following mean relative integrated squared estimation error, $\overline{\text{RISEE}} = \frac{1}{P} \sum_{p=1}^P \text{RISEE}_p$, is computed (for only FoFRM-QI, LQ, and RFoFRM-QI) to compare the estimation performance of the methods, \citep{Zhou21}, where
\begin{equation*}
\text{RISEE}_p = \text{Med} \lbrace \text{RISEE}_{p,j}; j = 1, \ldots, 250 \rbrace 
\end{equation*}
with
\begin{equation*}
\text{RISEE}_{p,j} = \frac{\Vert \beta_p(s,t) - \widehat{\beta}_p^{(j)}(s,t) \Vert_2^2}{\Vert \beta(s,t)_p\Vert_2^2}.
\end{equation*}
Here, $\widehat{\beta}_p^{(j)}(s,t)$ denotes the estimated parameter function obtained in the $j^\textsuperscript{th}$ replication and $\text{Med} \lbrace \cdot \rbrace$ is the median operator. In addition, for each sample size, $n_{\text{test}} = 100$ independent samples are generated as the test sample. With the generated test samples, the predictive performance of the methods is evaluated by applying the fitted models to the test samples, and the following mean squared prediction error (MSPE) is computed: 
\begin{equation*}
\text{MSPE} = \frac{1}{100} \sum_{i=1}^{100} \left\Vert \Y_i(t) - \widehat{\Y}_i(t) \right\Vert_2^2,
\end{equation*}
where $\widehat{\Y}_i(t)$ is the predicted response function for the $i^\textsuperscript{th}$ individual. In addition to comparing the methods' estimation and predictive performance, they are compared in terms of their outlier detection capabilities. For this purpose, the functional depth-based outlier detection method of \cite{Febrero08} (FGG) together with the $h$-modal depth proposed by \cite{cuevas07} is applied to the estimated residual functions $\widehat{\epsilon}_i(t) = \Y_i(t) - \widehat{\Y}_i(t)$ to determine the outliers in the response variable. This method makes it possible to determine both magnitude and shape outliers in the response variable \citep[see, e.g.,][for more details]{Harjit}. The area under the curve (AUC) of the receiver operating characteristic curve is computed to evaluate the outlier detection accuracy of the methods. 

For the LQ, FoFRM-QI, and RFoFRM-QI methods, the MSPE and AUC values are computed under four models:
\begin{inparaenum}
\item[1)] \textit{Full model}, where the FoFRM-QI is estimated using all main, quadratic, and interaction effect terms;
\item[2)] \textit{True model}, where the FoFRM-QI is estimated using only the relevant main, quadratic, and interaction effect terms;
\item[3)] \textit{Selected model}, where the FoFRM-QI is estimated using only the relevant main, quadratic, and interaction effect terms determined by the variable selection procedures; and
\item[4)] \textit{Main effect model}, where the FoFRM-QI is estimated using only all main effect terms.
\end{inparaenum}
These calculations aim to evaluate the performance of the variable selection procedure and compare the model fitting performance of FoFRM-QI and FoFRM. 

For the proposed method, the variable selection procedure discussed in the supplement is applied to the generated data to determine relevant main, quadratic, and interaction effect terms. For the FoFRM-QI, the usual BIC is used in the variable selection procedure. In contrast, for the LQ, the variable selection procedure presented by \cite{LuoQi} is used to determine relevant variables. The pffr and FDB methods are applied to the main effect model only. In the experiments, the RISEE values are only computed under the actual model to compare the estimation performance of the methods numerically. 

The numerical calculations are based on the cubic $B$-spline representation of the functional variables. For the main effect terms, 20 cubic $B$-spline basis functions are used to project the functional variables on the basis expansions. On the other hand, $20 \times 20 = 400$ tensor product cubic $B$-spline basis functions are used for the projections of quadratic and interaction effect terms. Furthermore, additional Monte-Carlo experiments are conducted to evaluate the asymptotic distribution of the proposed estimators ($\widehat{\beta}_{\tau p}(s,t)$, which is derived in the supplement as Normal distribution). In doing so, 250 Monte-Carlo experiments are conducted. In each experiment, the Anderson-Darling test for assessing multivariate normality \citep{mvnTest} is applied to $\widehat{\bm{B}}_p$s (the estimates computed in the finite-dimensional space) obtained by the proposed method. Then, we compute an average ratio of the estimates with normal distribution over 250 simulations (AD\%). Note that the truncation constants are set to $K_{\Y} = 2$ and $K_{\X_p} = 4$ when assessing the normality of the proposed estimators to obtain reasonable results for all the Monte-Carlo runs.

Throughout the experiments, $P = 6$ functional predictors are generated at 101 equally spaced points in the interval $[0, 1]$. We consider a similar process as in \cite{ivanescu2015} and \cite{Cai2020} to generate the functional predictors:
\begin{equation*}
\X_{ip}(s) = \sum_{r=1}^{10} \frac{1}{r^2} \left\lbrace \xi_{i1,r} \sqrt{2} \sin(r \pi s) + \xi_{i2,r} \sqrt{2} \cos(r \pi s) \right\rbrace,
\end{equation*}
where $\xi_{i1,r}$ and $\xi_{i2,r}$ for $r = 1, \ldots, 10$ are independently generated from a normal distribution with mean zero and variance 0.5. Then, the trajectories of the response variable are generated as follows:
\begin{align*}
\Y_i(t) &= \beta_0(t) + \sum_{p \in \mathcal{P}} \int_0^1 \X_{ip}(s) \beta_p(s,t) ds + \\
& \sum_{p \in \mathcal{M}} \sum_{p^{\prime} \in \mathcal{M}} \int_0^1 \int_0^1 \X_{ip}(r) \X_{ip^{\prime}}(s) \gamma_{p p^{\prime}}(r,s,t) dr ds + \epsilon_i(t),
\end{align*}
where $\mathcal{P} \in \left\lbrace 1, 2, 3, 4 \right\rbrace$ and $\mathcal{M} \in \left\lbrace ( 1,1 ), ( 1,4 ), ( 3,3 )
 (3,4) \right\rbrace$ respectively denote the index sets of relevant main and quadratic/interaction effect terms, the error terms $\epsilon_i(t) \sim N(0, \sigma_{\epsilon})$. Similar to \cite{ivanescu2015}, we consider two cases for the error terms; $\sigma_{\epsilon} = 0$ (noiseless case) and $\sigma_{\epsilon} = 0$ (noisy case). In the data generation process, the intercept function is set to $\beta_0(t) = 4 \cos(4 \pi t)$. The following two- and three-dimensional parameter functions are used to generate the trajectories of the response variable:
\begin{align*}
\beta_1(s,t) &= 2 \sin ( 2 \pi s ) + \sin ( \pi t ), \qquad \gamma_{11}(r,s,t) = e^{r^2 + s^2} \sqrt{t}, \\
\beta_2(s,t) &= 2 \cos ( \pi s ) + \cos (2 \pi t ), \qquad \gamma_{14}(r,s,t) = 2 \cos(\pi (r+s)) \sqrt{t}, \\
\beta_3(s,t) &= \cos ( \pi s ) + \sin (2 \pi t ), \qquad \gamma_{33}(r,s,t) = r + s + t^2 \\
\beta_4(s,t) &= 2 \sin ( 2 \pi s ) + \sin (2 \pi t ), \qquad \gamma_{44}(r,s,t) = (r^2 + s^2) t.
\end{align*}

Throughout the experiments, $n \times [5\%,~10\%]$ of the generated data (in the training sample) are contaminated by outliers to investigate the robustness of the proposed method. In doing so, we consider replacing randomly selected $n \times [5\%,~10\%]$ trajectories of the first two relevant functional predictors with $\widetilde{\X}(t)$ whose observations are generated as follows: 
\begin{equation*}
\widetilde{\X}_{ip}(s) = \sum_{r=1}^{10} \frac{1}{r^2} \left\lbrace \xi_{i1,r} \sqrt{6} \sin(r \pi s) + \xi_{i2,r} \sqrt{6} \cos(r \pi s) \right\rbrace.
\end{equation*}
In addition, we consider a contaminated normal distribution for the error terms, i.e., $\epsilon_i(t) \sim N(10, \sigma_{\epsilon})$ when generating the observations of the functional response. In this way, $n \times [5\%,~10\%]$ of the generated functional predictors are contaminated by shape outliers, while $n \times [5\%,~10\%]$ of the generated functional response are contaminated by magnitude outliers (see, e.g., \cite{BS19} for more details).

The results for the computed $\overline{\text{RISEE}}$ values for all the methods are presented in Table~\ref{tab:tab_1}. From the results, when no outlier is present in the data, the LQ produces considerably smaller $\overline{\text{RISEE}}$ values than the FoFRM-QI and proposed RFoFRM-QI methods. When outliers are present in the data, the LQ and FoFRM-QI are significantly affected by these observations and produce unsatisfactory parameter estimates, as demonstrated by their $\overline{\text{RISEE}}$ values. In this case, the parameter estimates are considerably larger than the ones they produce when no outlier exists in the data. On the other hand, the proposed method is not affected by outliers. It tends to produce similar $\overline{\text{RISEE}}$ values regardless of outliers and contamination levels with considerably less variability (median absolute deviation). 

The results presented in Table~\ref{tab:tab_1} also demonstrate that the LQ is affected by the noise level ($\sigma_{\epsilon}$). It produces larger $\overline{\text{RISEE}}$ in the noisy case compared with its $\overline{\text{RISEE}}$ values in the noiseless case. On the other hand, the FoFRM-QI and proposed method are unaffected by the noise level and generally produce similar $\overline{\text{RISEE}}$ for both noise levels. Furthermore, Table~\ref{tab:tab_1} shows that the estimates obtained by the proposed method generally follow a normal distribution so that the estimated parameters follow a normal distribution in at least 95\% over 250 Monte-Carlo runs.

\begin{center}
\begin{scriptsize}
\tabcolsep 0.13in
\renewcommand{\arraystretch}{0.86}
\begin{longtable}{@{}lllcccc@{}}\caption{\scriptsize{Computed $\overline{\text{RISEE}}$ values with their median absolute deviations (given in brackets) for the main effect parameters. $\overline{\text{RISEE}}$ values are computed for only LQ, FoFRM-QI, and RFoFRM-QI methods under the True model with two noise levels ($\sigma_{\epsilon} = [0, 0.5]$), three different contamination levels ($\text{CL} = [0\%,~5\%,~10\%]$), and three different sample sizes ($n = [100, 250, 500]$).}}\label{tab:tab_1}\\
\toprule
{$\sigma_{\epsilon}$} & {n} & {CL} & \multicolumn{3}{c}{$\overline{\text{RISEE}}$} & AD\% \\
\midrule
& & & LQ & FoFRM-QI & RFoFRM-QI \\
\cmidrule(l){4-4} \cmidrule(l){5-5} \cmidrule(l){6-6}
\multirow{9}{*}{0} & 100 & \multirow{3}{*}{0\%} & \textbf{0.0005} (0.0007) & 0.1987 (0.0286) & 0.2064 (0.0772) & 0.9550\\
& 250 & & \textbf{0.0004} (0.0004) & 0.1828 (0.0175) & 0.1069 (0.0741) & 0.9750 \\
& 500 & & \textbf{0.0004} (0.0003) & 0.0794 (0.0680) & 0.0607 (0.0337) & 0.9600 \\
\cmidrule(l){3-7}
& 100 & \multirow{3}{*}{5\%} & 0.6009 (0.2820) & 9.6249 (12.4372) & \textbf{0.2073} (0.0536) & 0.9650 \\
& 250 & & 0.2955 (0.1324) & 3.4218 (4.1281) & \textbf{0.1475} (0.0942) & 0.9750 \\
& 500 & & 0.2772 (0.1225) & 0.2505 (0.9755) & \textbf{0.0690} (0.0022) & 0.9750 \\
\cmidrule(l){3-7}
& 100 & \multirow{3}{*}{10\%} & 0.7778 (0.3412) & 15.7272 (17.3201) & \textbf{0.1961} (0.0784) & 0.9850 \\
& 250 & & 0.5368 (0.4865) & 3.2438 (4.0230) & \textbf{0.1791} (0.0618) & 0.9500 \\
& 500 & & 0.5993 (0.6431) & 0.6940 (0.8542) & \textbf{0.0671} (0.0078) & 0.9700 \\
\midrule
\multirow{9}{*}{0.5} & 100 & \multirow{3}{*}{0\%} & \textbf{0.0192} (0.0067) & 0.1898 (0.0235) & 0.1989 (0.0533) & 0.9650 \\
& 250 & & \textbf{0.0081} (0.0024) & 0.1865 (0.0131) & 0.1942 (0.0318) & 0.9750 \\
& 500 & & \textbf{0.0048} (0.0012) & 0.1813 (0.0106) & 0.1935 (0.0288) & 0.9850 \\
\cmidrule(l){3-7}
& 100 & \multirow{3}{*}{5\%} & 0.5387 (0.3105) & 14.5496 (13.4025) & \textbf{0.2192} (0.0695) & 0.9550 \\
& 250 & & 0.4857 (0.4441) & 5.4503 (6.4730) & \textbf{0.1889} (0.0283) & 0.9700 \\
& 500 & & 0.4298 (0.4188) & 0.9022 (1.8756) & \textbf{0.1891} (0.0336) & 0.9350 \\
\cmidrule(l){3-7}
& 100 & \multirow{3}{*}{10\%} & 0.6681 (0.2635) & 7.0049 (8.3756) & \textbf{0.2030} (0.0658) & 0.9800 \\
& 250 & & 0.5359 (0.4556) & 2.8610 (3.4932) & \textbf{0.1938} (0.0389) & 0.9850 \\
& 500 & & 1.2808 (0.9221) & 2.6767 (1.9963) & \textbf{0.1936} (0.0321) & 0.9850 \\
\bottomrule
\end{longtable}
\end{scriptsize}
\end{center}

\vspace{-.9in}

From Table~\ref{tab:tab_2}, when no outlier is present in the data, all the methods produce competitive predictive performance (i.e., smaller median MSPE values) when the quadratic and interaction effect terms are used in the model compared with the main effect model. The FoFRM-QI and proposed methods produce considerably larger MSPE values under the full model than their values obtained under the main effect model, i.e., their predictive performance is seriously affected by the redundant main and quadratic/interaction effect terms. In addition, this table shows that the FDB, pffr, LQ, and FoFRM-QI produce improved MSPE values than the proposed method when the generated data are not contaminated by the outliers. Moreover, in this case, the LQ produces similar or even better results under the true and selected models compared with its MSPE values obtained under the full model. On the other hand, the FoFRM-QI and proposed methods produce considerably smaller MSPE values under the true and selected models than their values obtained under the full model. In other words, when the redundant predictors and their quadratic/interaction effects are included in the model, the FoFRM-QI and proposed methods produce unsatisfactory predictive performance. However, they tend to produce reasonable MSPE values when only the relevant main effect terms and their quadratic/interaction effects are included in the model. It is evident from Table~\ref{tab:tab_2} that the variable selection procedures generally determine the relevant main, quadratic, and interaction effect terms for the models. 

When outliers are present in the data, the proposed method produces significantly smaller MSPE values with less variability than the FDB, pffr, LQ, and FoFRM-QI (except for full model, as observed under the outlier-free data case). The superiority of the proposed method over its competitors becomes more prominent as the contamination level increases. While the proposed method produces stable MSPE values under all the contamination levels, the LQ and FoFRM-QI tend to produce larger MSPE values as the contamination level increases. In addition, the LQ produces larger MSPE values under the selected model compared with its values obtained under the true model, i.e., outliers results in incorrect model selection for this method. On the other hand, the proposed method is unaffected by outliers and the variable selection procedure correctly determines the relevant main, quadratic, and interaction effect terms, leading to similar MSPE values under the true and selected models.

From Table~\ref{tab:tab_2}, all the methods generally produce similar MSPE values for different noise levels, $\sigma_{\epsilon}$. Sometimes, the methods produce smaller MSPE values under noisy data or vice versa. That is, the methods are not affected by the noise level for the considered data generation process. The pffr and LQ methods generally produce improved MSPE values than the FPCA-based methods. This is because the pffr and LQ methods apply an adaptive roughness penalty on the objective functions, leading to better control of the smoothness of the function estimates and improved results than the FPCA-based methods. In addition, the LQ method is based on the signal compression approaches, which provide a more effective basis expansion system for the coefficient functions than that of the FPCA-based methods \citep{Jin2023}.

Table~\ref{tab:tab_3} presents the results of the outlier detection capabilities of the methods. From this table, all the methods produce similar AUC values under the main effect model and they determine the outliers correctly with at least 97\% accuracy. Under the full model, while the LQ generally determines the outliers correctly, the FPCA-based models fail to produce satisfactory AUC values. The FPCA-based models do not fit the model adequately because of the redundant predictors and their quadratic/interaction effects. On the other hand, the proposed method produces significantly improved AUC values than the LQ and FoFRM-QI methods under both the true and selected models. The LQ and FoFRM-QI produce better AUC values under the main effect model compared with their results produced when the quadratic and interaction effect terms are included in the model. This may be because of the masking effect.

\begin{landscape}
\begin{center}
\begin{scriptsize}
\tabcolsep 0.02in
\renewcommand{\arraystretch}{0.92}
\begin{longtable}{@{}lllcccccccccccccc@{}}\caption{\scriptsize{Computed median MSPE values with their median absolute deviations (given in brackets). MSPE values are computed for the FDB, pffr, LQ, FoFRM-QI, and RFoFRM-QI methods under the \textit{Main effect model} (Main), \textit{Full model} (Full), \textit{True model} (True), and \textit{Selected model} (Selected) with two noise levels ($\sigma_{\epsilon} = [0, 0.5]$), three different contamination levels ($\text{CL} = [0\%,~5\%,~10\%]$), and three different sample sizes ($n = [100, 250, 500]$).}}\label{tab:tab_2}\\
\toprule
{$\sigma_{\epsilon}$} & {$n$} & {CL} & \multicolumn{14}{c}{Model} \\
& & & \multicolumn{5}{c}{Main} & \multicolumn{3}{c}{Full}  & \multicolumn{3}{c}{True} & \multicolumn{3}{c}{Selected}\\
\cmidrule(l){4-8} \cmidrule(l){9-11} \cmidrule(l){12-14} \cmidrule(l){15-17}
& & & FDB & pffr & LQ & FoFRM-QI & RFoFRM-QI & LQ & FoFRM-QI & RFoFRM-QI & LQ & FoFRM-QI & RFoFRM-QI & LQ & FoFRM-QI & RFoFRM-QI \\
\cmidrule(l){4-17}
\multirow{18}{*}{1} & 100 & \multirow{6}{*}{0\%} & 0.3860 & 0.4525 & \textbf{0.3323} & 0.4122 & 0.4160 & \textbf{0.0077} & 1.3957 & 2.4771 & \textbf{0.0010} & 0.0617 & 0.1443 & \textbf{0.0010} & 0.0744 & 0.2668 \\
& & & (0.1431) & (0.0815) & (0.0925) & (0.1127) & (0.1101) & (0.0024) & (0.5015) & (0.3864) & (0.0001) & (0.0075) & (0.0821) & (0.0001) & (0.0195) & (0.1589) \\
& 250 & & 0.3855 & 0.3553 & \textbf{0.3428} & 0.3662 & 0.3530 & \textbf{0.0009} & 0.0839 & 0.2719 & \textbf{0.0009} & 0.0516 & 0.0508 & \textbf{0.0009} & 0.0724 & 0.0767 \\
& & & (0.0737) & (0.0949) & (0.0768) & (0.0775) & (0.0882) & (0.0001) & (0.0189) & (0.1094) & (0.0001) & (0.0106) & (0.0278) & (0.0001) & (0.0321) & (0.0445) \\
& 500 & & 0.3871 & 0.3353 & \textbf{0.3293} & 0.3346 & 0.3311 & \textbf{0.0010} & 0.0174 & 0.1086 & \textbf{0.0010} & 0.0170 & 0.0317 & \textbf{0.0010} & 0.0483 & 0.0399 \\
& & & (0.0788) & (0.0685) & (0.0725) & (0.0740) & (0.0777) & (0.0002) & (0.0026) & (0.0300) & (0.0002) & (0.0026) & (0.0155) & (0.0002) & (0.0361) & (0.0238) \\
\cmidrule(l){4-8} \cmidrule(l){9-11} \cmidrule(l){12-14} \cmidrule(l){15-17}
& 100 & \multirow{6}{*}{5\%} & 1.1561 & 2.7379 & 1.584 & 2.9463 & \textbf{0.3989} & \textbf{1.4766} & 4.7016 & 4.8922 & 1.468 & 8.3720 & \textbf{1.0691} & 2.050 & 2.4128 & \textbf{0.8401} \\
& & & (0.6881) & (0.3979) & (0.3876) & (0.9575) & (0.0729) & (0.5088) & (2.4340) & (2.4063) & (0.5705) & (8.0826) & (1.4064) & (1.1029) & (0.1984) & (0.9442) \\
& 250 & & 0.8463 & 1.3138 & 0.8850 & 1.6545 & \textbf{0.3453} & \textbf{1.2054} & 5.8124 & 4.2429 & 0.9281 & 2.7428 & \textbf{0.0896} & 1.2132 & 2.0860 & \textbf{0.1034} \\
& & & (0.2780) & (0.2063) & (0.2334) & (0.2826) & (0.0783) & (0.2372) & (2.4779) & (1.7881) & (0.3137) & (1.3331) & (0.0461) & (0.4617) & (0.2544) & (0.1080) \\
& 500 & & 0.8057 & 0.8027 & 0.6658 & 1.1915 & \textbf{0.3260} & \textbf{1.1196} & 7.1321 & 3.5679 & 0.9770 & 1.1048 & \textbf{0.0629} & 1.4943 & 2.3005 & \textbf{0.0616} \\
& & & (0.4457) & (0.4389) & (0.3117) & (0.9650) & (0.1109) & (0.9985) & (2.4475) & (1.0011) & (0.3321) & (0.9977) & (0.0301) & (0.8745) & (1.0078) & (0.0324) \\
\cmidrule(l){4-8} \cmidrule(l){9-11} \cmidrule(l){12-14} \cmidrule(l){15-17}
& 100 & \multirow{6}{*}{10\%} & 2.2290 & 5.5170 & 2.7419 & 5.7987 & \textbf{0.3952} & \textbf{2.7486} & 8.3677 & 7.2952 & 2.5755 & 21.4602 & \textbf{0.9809} & 3.6925 & 3.1849 & \textbf{1.0057} \\
& & & (1.6262) & (0.4323) & (0.4385) & (1.7878) & (0.1043) & (0.6769) & (3.6259) & (2.3524) & (0.9204) & (10.5152) & (1.1248) & (1.7368) & (0.8629) & (1.6286) \\
& 250 & & 1.6897 & 2.6982 & 1.8653 & 2.9604 & \textbf{0.4027} & \textbf{2.2486} & 9.9174 & 6.9191 & 2.3099 & 3.6546 & \textbf{0.1955} & 2.2939 & 3.0440 & \textbf{0.2136} \\
& & & (0.5249) & (0.2848) & (0.4599) & (0.6074) & (0.1022) & (0.4522) & (3.8199) & (4.0790) & (0.6906) & (1.7862) & (0.1000) & (0.5536) & (0.5558) & (0.2460) \\
& 500 & & 1.0939 & 1.5863 & 1.3664 & 1.6972 & \textbf{0.4720} & \textbf{3.5666} & 5.1432 & 5.0199 & 2.4298 & 2.9019 & \textbf{0.1421} & 2.8335 & 2.4336 & \textbf{0.1298} \\
& & & (0.4988) & (0.2478) & (0.4011) & (0.5074) & (0.0900) & (0.4020) & (2.2214) & (3.1010) & (0.1431) & (0.0925) & (0.0801) & (0.0908) & (0.1398) & (0.1088) \\
\midrule
{$\sigma_{\epsilon}$} & {$n$} & {CL} & \multicolumn{14}{c}{Model} \\
& & & \multicolumn{5}{c}{Main} & \multicolumn{3}{c}{Full}  & \multicolumn{3}{c}{True} & \multicolumn{3}{c}{Selected}\\
\cmidrule(l){4-8} \cmidrule(l){9-11} \cmidrule(l){12-14} \cmidrule(l){15-17}
& & & FDB & pffr & LQ & FoFRM-QI & RFoFRM-QI & LQ & FoFRM-QI & RFoFRM-QI & LQ & FoFRM-QI & RFoFRM-QI & LQ & FoFRM-QI & RFoFRM-QI \\
\cmidrule(l){4-17}
\multirow{18}{*}{2} & 100 & \multirow{6}{*}{0\%} & 0.3424 & 0.3553 & \textbf{0.2989} & 0.3346 & 0.3247 & \textbf{0.0387} & 2.50361 & 2.8630 & \textbf{0.0081} & 0.0611 & 0.1872 & \textbf{0.0132} & 0.0988 & 0.2381 \\
& & & (0.0770) & (0.0933) & (0.0848) & (0.0865) & (0.0793) & (0.0083) & (1.2107) & (0.5829) & (0.0006) & (0.0104) & (0.1575) & (0.0010) & (0.0253) & (0.0660) \\
& 250 & & 0.3503 & 0.3243 & \textbf{0.3131} & 0.3364 & 0.3403 & \textbf{0.0085} & 0.0821 & 0.1585 & \textbf{0.0032} & 0.0532 & 0.0767 & \textbf{0.0042} & 0.1205 & 0.1273 \\
& & & (0.0537) & (0.0690) & (0.0585) & (0.0598) & (0.0679) & (0.0009) & (0.0118) & (0.0607) & (0.0003) & (0.0068) & (0.0125) & (0.0005) & (0.0802) & (0.0774) \\
& 500 & & 0.3553 & 0.3337 & \textbf{0.3281} & 0.3505 & 0.3661 & \textbf{0.0038} & 0.0608 & 0.0765 & \textbf{0.0023} & 0.0539 & 0.0687 & \textbf{0.0026} & 0.1100 & 0.1164 \\
& & & (0.1148) & (0.0727) & (0.1045) & (0.0864) & (0.1076) & (0.0003) & (0.0089) & (0.0098) & (0.0001) & (0.0048) & (0.0056) & (0.0001) & (0.0199) & (0.0469) \\
\cmidrule(l){4-8} \cmidrule(l){9-11} \cmidrule(l){12-14} \cmidrule(l){15-17}
& 100 & \multirow{6}{*}{5\%} & 1.1786 & 2.8502 & 1.3882 & 3.0537 & \textbf{0.3996} & \textbf{1.5303} & 5.6728 & 5.4385 & 2.2538 & 16.9662 & \textbf{0.2040} & 2.1076 & 2.4983 & \textbf{0.2715} \\
& & & (0.7807) & (0.1956) & (0.2595) & (0.7054) & (0.0706) & (0.4665) & (2.0391) & (3.7222) & (1.4461) & (9.4836) & (0.1663) & (1.4056) & (0.4885) & (0.0812) \\
& 250 & & 0.7829 & 1.2662 & 0.7460 & 1.5945 & \textbf{0.3161} & \textbf{1.2013} & 5.3424 & 1.5618 & 0.8367 & 3.2264 & \textbf{0.1020} & 1.3862 & 2.1791 & \textbf{0.1436} \\
& & & (0.2354) & (0.1764) & (0.1324) & (0.1745) & (0.0588) & (0.1971) & (2.0030) & (1.5122) & (0.3652) & (1.0411) & (0.0302) & (0.3991) & (0.4602) & (0.0506) \\
& 500 & & 0.5055 & 1.0267 & 0.7963 & 1.0212 & \textbf{0.2893} & \textbf{1.0211} & 2.8368 & 0.1025 & 0.6389 & 0.8082 & \textbf{0.1038} & 0.8964 & 2.0236 & \textbf{0.1592} \\
& & & (0.6878) & (0.9144) & (0.6001) & (0.9974) & (0.1101) & (0.9902) & (1.0587) & (0.0965) & (0.7014) & (0.8878) & (0.0730) & (0.6780) & (1.0244) & (0.1101) \\
\cmidrule(l){4-8} \cmidrule(l){9-11} \cmidrule(l){12-14} \cmidrule(l){15-17}
& 100 & \multirow{6}{*}{10\%} & 2.4312 & 4.0448 & 3.0012 & 4.5853 & \textbf{0.4590} & \textbf{3.2268} & 8.8526 & 8.6673 & 2.8557 & 11.7713 & \textbf{0.7345} & 3.5516 & 3.3931 & \textbf{0.8310} \\
& & & (0.6108) & (0.3794) & (0.3113) & (0.4722) & (0.1564) & (0.9103) & (3.5174) & (2.0998) & (0.6292) & (12.3383) & (0.8335) & (1.1325) & (0.8092) & (0.6862) \\
& 250 & & 1.7025 & 2.4763 & 1.8368 & 2.7749 & \textbf{0.4068} & \textbf{2.2030} & 11.1600 & 3.7436& 2.0995 & 3.8149 & \textbf{0.1738} & 2.5480 & 2.8322 & \textbf{0.1693} \\
& & & (0.4740) & (0.2109) & (0.4033) & (0.4709) & (0.1090) & (0.5808) & (4.4515) & (3.0343) & (0.5197) & (1.6242) & (0.0498) & (0.5602) & (0.4637) & (0.1128) \\
& 500 & & 1.9682 & 2.0335 & 1.4606 & 2.2186 & \textbf{0.3568} & 1.6225 & 9.4438 & \textbf{0.1492} & 1.5506 & 2.4439 & \textbf{0.1470} & 1.2801 & 3.1836 & \textbf{0.1648} \\
& & & (0.4117) & (0.6879) & (0.1544) & (0.2114) & (0.0755) & (0.2111) & (2.0114) & (0.1052) & (0.8447) & (1.0335) & (0.1002) & (0.9544) & (1.2552) & (0.1190) \\
\bottomrule
\end{longtable}
\end{scriptsize}
\end{center}
\end{landscape}

\begin{landscape}
\begin{center}
\begin{scriptsize}
\tabcolsep 0.02in
\renewcommand{\arraystretch}{0.92}
\begin{longtable}{@{}lllcccccccccccccc@{}}\caption{\scriptsize{Computed median AUC values. AUC values are computed for the FDB, pffr, LQ, FoFRM-QI, and RFoFRM-QI methods under the \textit{Main effect model} (Main), \textit{Full model} (Full), \textit{True model} (True), and \textit{Selected model} (Selected) with two noise levels ($\sigma_{\epsilon} = [0, 0.5]$), three different contamination levels ($\text{CL} = [0\%,~5\%,~10\%]$), and three different sample sizes ($n = [100, 250, 500]$).}}\label{tab:tab_3}\\
\toprule
{$\sigma_{\epsilon}$} & {$n$} & {CL} & \multicolumn{14}{c}{Model} \\
& & & \multicolumn{5}{c}{Main} & \multicolumn{3}{c}{Full}  & \multicolumn{3}{c}{True} & \multicolumn{3}{c}{Selected}\\
\cmidrule(l){4-8} \cmidrule(l){9-11} \cmidrule(l){12-14} \cmidrule(l){15-17}
& & & FDB & pffr & LQ & FoFRM-QI & RFoFRM-QI & LQ & FoFRM-QI & RFoFRM-QI & LQ & FoFRM-QI & RFoFRM-QI & LQ & FoFRM-QI & RFoFRM-QI \\
\cmidrule(l){4-17}
\multirow{9}{*}{1} & 100 & \multirow{3}{*}{5\%} & 0.9894 & 0.9915 & \textbf{1.0000} & 0.9673 & 0.9936 & \textbf{0.91473} & 0.3789 & 0.7842 & 0.9510 & 0.5768 & \textbf{0.9652} & 0.8915 & 0.9226 & \textbf{0.9910} \\
& 250 & & 0.9901 & 0.9926 & 0.9908 & \textbf{0.9950} & 0.9747 & \textbf{0.9457} & 0.3960 & 0.6911 & 0.9556 & 0.8574 & \textbf{0.9933} & 0.8911 & 0.9575 & \textbf{0.9896} \\
& 500 & & 0.9849 & 0.9898 & 0.9860 & \textbf{0.9942} & 0.9845 & \textbf{0.9829} & 0.6576 & 0.7110 & 0.9713 & 0.9673 & \textbf{0.9978} & 0.9270 & 0.9544 & \textbf{0.9986} \\
\cmidrule(l){4-8} \cmidrule(l){9-11} \cmidrule(l){12-14} \cmidrule(l){15-17}
& 100 & \multirow{3}{*}{10\%} & 0.9844 & 0.9827 & \textbf{0.9894} & 0.9855 & 0.9766 & \textbf{0.8555} & 0.3405 & 0.6933 & 0.9472 & 0.4411 & \textbf{0.9733} & 0.8805 & 0.9066 & \textbf{0.9688} \\
& 250 & & 0.9776 & 0.9843 & 0.9817 & \textbf{0.9911} & 0.9766 & \textbf{0.91946} & 0.3528 & 0.6929 & 0.9117 & 0.8482 & \textbf{0.9836} & 0.8807 & 0.9167 & \textbf{0.9887} \\
& 500 & & 0.9695 & 0.9679 & 0.9633 & \textbf{0.9749} & 0.9604 & 0.84310 & \textbf{0.9206} & 0.8192 & 0.9093 & 0.9272 & \textbf{0.9890} & 0.8820 & 0.9385 & \textbf{0.9890} \\
\midrule
\multirow{9}{*}{2} & 100 & \multirow{3}{*}{5\%} & \textbf{1.0000} & \textbf{1.0000} & \textbf{1.0000} & \textbf{1.0000} & 0.9936 & \textbf{0.8621} & 0.3178 & 0.7431 & 0.9715 & 0.3831 & \textbf{1.0000} & 0.9636 & 0.9326 & \textbf{1.0000} \\
& 250 & & 0.9950 & 0.9984 & 0.9957 & \textbf{1.0000} & 0.9959 & \textbf{0.9453} & 0.3737 & 0.8839 & 0.9537 & 0.8741 & \textbf{1.0000} & 0.8595 & 0.9318 & \textbf{1.0000} \\
& 500 & & \textbf{1.0000} & 0.9967 & 0.9985 & \textbf{1.0000} & 0.9978 & 0.8661 & 0.7793 & \textbf{1.0000} & 0.9674 & 0.9670 & \textbf{1.0000} & 0.9586 & 0.9577 & \textbf{0.9980} \\
\cmidrule(l){4-8} \cmidrule(l){9-11} \cmidrule(l){12-14} \cmidrule(l){15-17}
& 100 & \multirow{3}{*}{10\%} & 0.9922 & 0.9711 & 0.9944 & 0.9711 & \textbf{0.9988} & \textbf{0.9155} & 0.3233 & 0.6966 & 0.9366 & 0.5788 & \textbf{0.9811} & 0.9022 & 0.8722 & \textbf{0.9888} \\
& 250 & & 0.9917 & 0.9930 & 0.9922 & \textbf{0.9954} & 0.9911 & \textbf{0.9210} & 0.3709 & 0.7944 & 0.9210 & 0.8988 & \textbf{1.0000} & 0.7994 & 0.9171 & \textbf{0.9981} \\
& 500 & & 0.9864 & 0.9918 & 0.9858 & \textbf{0.9934} & 0.9912 & 0.9417 & 0.6115 & \textbf{1.0000} & 0.9290 & 0.8952 & \textbf{1.0000} & 0.9547 & 0.9494 & \textbf{0.9991} \\
\bottomrule
\end{longtable}
\end{scriptsize}
\end{center}
\end{landscape}

\section{Conclusion} \label{sec:conc}

The FoFRM has become a general framework for investigating a linear relationship between a functional response and a set of functional predictors. In this model, the relationship between the functional response and a functional predictor is assumed to be independent of other functional predictors. However, this independence assumption may need to be more relaxed in empirical applications because the predictors may interact. The FoFRM-QI has been proposed to overcome this problem, and several strategies have been developed to estimate this model. 

We propose a novel method to estimate the FoFRM-QI robustly. The existing estimation strategies, such as basis expansion, FPCA, and functional partial least squares, use non-robust estimators. In the presence of outliers, the non-robust methods may produce poor model fitting because the effects of outliers in the quadratic and interaction effect terms may deteriorate the model structure more severely than their main effect. The proposed method consists of two steps:
\begin{inparaenum}
\item[1)] The functional objects in the model are decomposed via an RFPCA. The RFPCA approximates the infinite-dimensional model in the finite-dimensional space of FPCA basis functions. 
\item[2)] The parameters of the approximate model are estimated using the $\tau$-estimator. An RBIC determines the optimum number of principal components to make the proposed method more practical. In addition, a forward stepwise variable selection procedure is used to determine the relevant main, quadratic, and interaction effect terms and improve the model's predictive performance. 
\end{inparaenum}
The proposed method's asymptotic consistency and influence function are investigated under mild conditions. 

Via several Monte-Carlo experiments, the proposed method produces improved estimation and predictive performance over the existing methods in the presence of outliers. It produces competitive results to the available methods when no outlier exists in the data. The proposed method is applied to the U.S. COVID-19 data to robustly and effectively explore the dynamic of COVID-19 prevalence in the U.S. Our results demonstrate that using quadratic and interaction effect terms enables us to understand better the changes in the response variable (total number of deaths). In addition, the results show that the proposed method generally produces better model fitting for the U.S. COVID-19 data compared with the existing non-robust methods. 

There are several possible directions where the proposed method can be extended further.
\begin{inparaenum}
\item[1)] From the point of view of empirical data example, the proposed method can be used to explore the dynamic of COVID-19-related quantities, such as the total number of confirmed cases and the number of COVID-19 patients in the ICU.
\item[2)] In the present paper, only the functional predictors are considered in the model. However, the dynamics of COVID-19 data may depend on non-functional (scalar) variables, such as internal movement restrictions, mobility, mitigation, and policy changes. The proposed method can be extended to the case where both the functional and scalar predictors are used in the model to explain the COVID-19 dynamics better.
\item[3)] The proposed method works for only densely observed functional data. However, sparse functional data, where only a few irregularly-spaced measurements are available for each subject, are common in many real-world applications \citep[see, e.g.,][]{Qiyao2021}. The proposed method can also be extended to sparse functional data by combining the RFPCA and the robust sparse principal component analysis proposed by \cite{croux13}.
\end{inparaenum}

\section*{Acknowledgment}

We would like to thank two reviewers for their careful reading of our manuscript and valuable suggestions and comments, which have helped us produce an improved version of our manuscript. The first author was supported by The Scientific and Technological Research Council of Turkey (TUBITAK) (grant no: 120F270).

\bibliography{rfq}

\begin{thebibliography}{43}
\providecommand{\natexlab}[1]{#1}
\providecommand{\url}[1]{\texttt{#1}}
\expandafter\ifx\csname urlstyle\endcsname\relax
  \providecommand{\doi}[1]{doi: #1}\else
  \providecommand{\doi}{doi: \begingroup \urlstyle{rm}\Url}\fi

\bibitem[Agull\'o et~al.(2008)Agull\'o, Croux, and Aelst]{Agullo2008}
{\rm Agull\'o, J., Croux, C., {\rm and} Aelst, S.~V.} (2008).
\newblock {The multivariate least-trimmed squares estimator}.
\newblock \emph{Journal of Multivariate Analysis}, {\bf 99}\penalty0 (3),
  \penalty0 311--338.

\bibitem[Bali et~al.(2011)Bali, Boente, Tyler, and Wang]{Bali2011}
{\rm Bali, J.~L., Boente, G., Tyler, D.~E., {\rm and} Wang, J.-L.} (2011).
\newblock {Robust functional principal components: A projection-pursuit
  approach}.
\newblock \emph{The Annals of Statistics}, {\bf 39}\penalty0 (6), \penalty0
  2852--2882.

\bibitem[Beaton and Tukey(1974)]{Beaton1974}
{\rm Beaton, A.~E. {\rm and} Tukey, J.~W.} (1974).
\newblock {The fitting of power series, meaning polynomials, illustrated on
  band-spectroscopic data}.
\newblock \emph{Technometrics}, {\bf 16}\penalty0 (2), \penalty0 147--185.

\bibitem[Ben et~al.(2006)Ben, Martinez, and Yohai]{Ben2006}
{\rm Ben, M.~G., Martinez, E., {\rm and} Yohai, V.~J.} (2006).
\newblock {Robust estimation for the multivariate linear model based on a
  $\tau$ scale}.
\newblock \emph{Journal of Multivariate Analysis}, {\bf 97}\penalty0 (7),
  \penalty0 1600--1622.

\bibitem[Beyaztas and Shang(2019)]{BS19}
{\rm Beyaztas, U. {\rm and} Shang, H.~L.} (2019).
\newblock {Forecasting functional time series using weighted likelihood
  methodology}.
\newblock \emph{Journal of Statistical Computation and Simulation}, {\bf
  89}\penalty0 (16), \penalty0 3046--3060.

\bibitem[Beyaztas and Shang(2020)]{BS20}
{\rm Beyaztas, U. {\rm and} Shang, H.~L.} (2020).
\newblock {On function-on-function regression: Partial least squares approach}.
\newblock \emph{Environmental and Ecological Statistics}, {\bf 27}\penalty0
  (1), \penalty0 95--114.

\bibitem[Beyaztas and Shang(2021)]{BS22}
{\rm Beyaztas, U. {\rm and} Shang, H.~L.} (2021).
\newblock {A partial least squares approach for function-on-function
  interaction regression}.
\newblock \emph{Computational Statistics}, {\bf 36}\penalty0 (2), \penalty0
  911--939.

\bibitem[Beyaztas and Shang(2022{\natexlab{a}})]{BS2022BJPS}
{\rm Beyaztas, U. {\rm and} Shang, H.~L.} (2022{\natexlab{a}}).
\newblock {A robust partial least squares approach for function-on-function
  regression}.
\newblock \emph{Brazilian Journal of Probability and Statistics}, {\bf
  36}\penalty0 (2), \penalty0 199--219.

\bibitem[Beyaztas and Shang(2022{\natexlab{b}})]{BS20C}
{\rm Beyaztas, U. {\rm and} Shang, H.~L.} (2022{\natexlab{b}}).
\newblock {A comparison of parameter estimation in function-on-function
  regression}.
\newblock \emph{Communications in Statistics -- Simulation and Computation},
  {\bf 51}\penalty0 (2), \penalty0 4607--4637.

\bibitem[Bilodeau and Duchesne(2000)]{Bilodeau2000}
{\rm Bilodeau, M. {\rm and} Duchesne, P.} (2000).
\newblock {Robust estimation of the SUR model}.
\newblock \emph{The Canadian Journal of Statistics}, {\bf 28}\penalty0 (2),
  \penalty0 277--288.

\bibitem[Brockhaus et~al.(2020)Brockhaus, R\"ugamer, and Greven]{FDboost}
{\rm Brockhaus, S., R\"ugamer, D., {\rm and} Greven, S.} (2020).
\newblock Boosting functional regression models with {FDboost}.
\newblock \emph{Journal of Statistical Software}, {\bf 94}\penalty0 (10),
  \penalty0 1--50.

\bibitem[Cai et~al.(2021)Cai, Xue, and Cao]{Cai2020}
{\rm Cai, X., Xue, L., {\rm and} Cao, J.} (2021).
\newblock {Robust penalized {M}-estimation for function-on-function linear
  regression}.
\newblock \emph{Stat}, {\bf 10}, \penalty0 e390.

\bibitem[Cai et~al.(2022)Cai, Xue, and Cao]{Cai2022}
{\rm Cai, X., Xue, L., {\rm and} Cao, J.} (2022).
\newblock {Variable selection for multiple function-on-function linear
  regression}.
\newblock \emph{Statistica Snica}, {\bf 1032}\penalty0 (3), \penalty0
  1435--1465.

\bibitem[Chiou et~al.(2016)Chiou, Yang, and Chen]{chiou2016}
{\rm Chiou, J.-M., Yang, Y.-F., {\rm and} Chen, Y.-T.} (2016).
\newblock Multivariate functional linear regression and prediction.
\newblock \emph{Journal of Multivariate Analysis}, {\bf 146}, \penalty0
  301--312.

\bibitem[Croux and Ruiz-Gazen(1996)]{croux96}
{\rm Croux, C. {\rm and} Ruiz-Gazen, A.} (1996).
\newblock High breakdown estimators for principal components: the
  projection-pursuit approach revisited.
\newblock \emph{Journal of Multivariate Analysis}, {\bf 95}\penalty0 (1),
  \penalty0 206--226.

\bibitem[Croux et~al.(2013)Croux, Filzmoser, and Fritz]{croux13}
{\rm Croux, C., Filzmoser, P., {\rm and} Fritz, H.} (2013).
\newblock Robust sparse principal component analysis.
\newblock \emph{Technometrics}, {\bf 55}\penalty0 (2), \penalty0 202--214.

\bibitem[Cuaves et~al.(2007)Cuaves, Febrero, and Fraiman]{cuevas07}
{\rm Cuaves, A., Febrero, M., {\rm and} Fraiman, R.} (2007).
\newblock Robust estimation and classification for functional data via
  projection-based depth notions.
\newblock \emph{Computational Statistics}, {\bf 22}\penalty0 (3), \penalty0
  481--496.

\bibitem[de~Lathuauwer et~al.(2000)de~Lathuauwer, de~Moor, and
  Vandewalle]{Lathuawer}
{\rm de~Lathuauwer, L., de~Moor, B., {\rm and} Vandewalle, J.} (2000).
\newblock A multilinear singular value decomposition.
\newblock \emph{SIAM Journal on Matrix Analysis and Applications}, {\bf
  21}\penalty0 (4), \penalty0 1253--1278.

\bibitem[Dutta and Genton(2017)]{Dutta2017}
{\rm Dutta, S. {\rm and} Genton, M.~G.} (2017).
\newblock Depth-weighted robust multivariate regression with application to
  sparse data.
\newblock \emph{The Canadian Journal of Statistics}, {\bf 45}\penalty0 (2),
  \penalty0 164--184.

\bibitem[Febrero-Bande et~al.(2008)Febrero-Bande, Galeano, and
  Gonzalez-Mantelga]{Febrero08}
{\rm Febrero-Bande, M., Galeano, P., {\rm and} Gonzalez-Mantelga, W.} (2008).
\newblock {Outlier detection in functional data by depth measures, with
  application to identify abnormal NO$_x$ levels}.
\newblock \emph{Environmetrics}, {\bf 19}\penalty0 (4), \penalty0 331--345.

\bibitem[Greven and Scheipl(2017)]{Greven2017}
{\rm Greven, S. {\rm and} Scheipl, F.} (2017).
\newblock A general framework for functional regression modelling.
\newblock \emph{Statistical Modelling}, {\bf 17}\penalty0 (1-2), \penalty0
  1--35.

\bibitem[H\"{o}ssjer(1992)]{Hosser1992}
{\rm H\"{o}ssjer, O.} (1992).
\newblock {On the optimality of S-estimators}.
\newblock \emph{Statistics \& Probability Letters}, {\bf 14}\penalty0 (5),
  \penalty0 413--419.

\bibitem[Hullait et~al.(2021)Hullait, Leslie, Pavlidis, and King]{Harjit}
{\rm Hullait, H., Leslie, D.~S., Pavlidis, N.~G., {\rm and} King, S.} (2021).
\newblock Robust function-on-function regression.
\newblock \emph{Technometrics}, {\bf 63}\penalty0 (3), \penalty0 396--409.

\bibitem[Ivanescu et~al.(2015)Ivanescu, Staicu, Scheipl, and
  Greven]{ivanescu2015}
{\rm Ivanescu, A.~E., Staicu, A.-M., Scheipl, F., {\rm and} Greven, S.} (2015).
\newblock Penalized function-on-function regression.
\newblock \emph{Computational Statistics}, {\bf 30}\penalty0 (2), \penalty0
  539--568.

\bibitem[Jin et~al.(2023)Jin, Sun, and Du]{Jin2023}
{\rm Jin, H., Sun, X., {\rm and} Du, P.} (2023).
\newblock Optimal function-on-function regression with interaction between
  functional predictors.
\newblock \emph{Statistica Sinica}, {\bf in press}.

\bibitem[Kalogridis and Aelst(2019)]{kalogridis2019}
{\rm Kalogridis, I. {\rm and} Aelst, S.~V.} (2019).
\newblock Robust functional regression based on principal components.
\newblock \emph{Journal of Multivariate Analysis}, {\bf 173}, \penalty0
  393--415.

\bibitem[Koenker and Portnoy(1990)]{Koenker1990}
{\rm Koenker, R. {\rm and} Portnoy, S.} (1990).
\newblock M estimation of multivariate regressions.
\newblock \emph{Journal of the American Statistical Association: Theory and
  Methods}, {\bf 85}\penalty0 (412), \penalty0 1060--1068.

\bibitem[Kudraszow and Moronna(2011)]{kudraszow2011}
{\rm Kudraszow, N.~L. {\rm and} Moronna, R.~A.} (2011).
\newblock Estimates of {MM} type for the multivariate linear model.
\newblock \emph{Journal of Multivariate Analysis}, {\bf 102}\penalty0 (9),
  \penalty0 1280--1292.

\bibitem[Luo and Qi(2019)]{LuoQi}
{\rm Luo, R. {\rm and} Qi, X.} (2019).
\newblock Interaction model and model selection for function-on-function
  regression.
\newblock \emph{Journal of Computational and Graphical Statistics}, {\bf
  28}\penalty0 (2), \penalty0 309--322.

\bibitem[Luo and Qi(2022)]{LuoQi21}
{\rm Luo, R. {\rm and} Qi, X.} (2022).
\newblock {Restricted function‐on‐function linear regression model}.
\newblock \emph{Biometrics}, {\bf 78}\penalty0 (3), \penalty0 1031--1044.

\bibitem[Matsui(2020)]{Matsui2020}
{\rm Matsui, H.} (2020).
\newblock Quadratic regression for functional response models.
\newblock \emph{Econometrics and Statistics}, {\bf 13}, \penalty0 125--136.

\bibitem[M{\"u}ller and Yao(2008)]{MullerYao2008}
{\rm M{\"u}ller, H.-G. {\rm and} Yao, F.} (2008).
\newblock Functional additive models.
\newblock \emph{Journal of the American Statistical Association: Theory and
  Methods}, {\bf 103}\penalty0 (484), \penalty0 1534--1544.

\bibitem[Pya et~al.(2016)Pya, Voinov, Makarov, and Voinov]{mvnTest}
{\rm Pya, N., Voinov, V., Makarov, R., {\rm and} Voinov, Y.} (2016).
\newblock \emph{mvnTest: Goodness of Fit Tests for Multivariate Normality}.
\newblock URL \url{https://CRAN.R-project.org/package=mvnTest}.
\newblock R package version 1.1-0.

\bibitem[Ramsay and Dalzell(1991)]{ramsay1991}
{\rm Ramsay, J.~O. {\rm and} Dalzell, C.~J.} (1991).
\newblock Some tools for functional data analysis.
\newblock \emph{Journal of the Royal Statistical Society, Series B}, {\bf
  53}\penalty0 (3), \penalty0 539--572.

\bibitem[Rousseeuw et~al.(2004)Rousseeuw, Driessen, Aelst, and
  Agull\'o]{Rousseeuw2004}
{\rm Rousseeuw, P.~J., Driessen, K.~V., Aelst, S.~V., {\rm and} Agull\'o, J.}
  (2004).
\newblock Robust multivariate regression.
\newblock \emph{Technometrics}, {\bf 46}\penalty0 (3), \penalty0 293--305.

\bibitem[Sun and Wang(2020)]{SunWang}
{\rm Sun, Y. {\rm and} Wang, Q.} (2020).
\newblock Function-on-function quadratic regression models.
\newblock \emph{Computational Statistics \& Data Analysis}, {\bf 142},
  \penalty0 106814.

\bibitem[Wang(2021)]{Qiyao2021}
{\rm Wang, Q.} (2021).
\newblock Two-sample inference for sparse functional data.
\newblock \emph{Electronic Journal of Statistics}, {\bf 15}, \penalty0
  1395--1423.

\bibitem[Wang(2014)]{wang2014}
{\rm Wang, W.} (2014).
\newblock Linear mixed function‐on‐function regression models.
\newblock \emph{Biometrics}, {\bf 70}\penalty0 (4), \penalty0 794--801.

\bibitem[Wang et~al.(2022)Wang, Dong, Ma, and Wang]{Wang2022}
{\rm Wang, Z., Dong, H., Ma, P., {\rm and} Wang, T.} (2022).
\newblock Estimation and model selection for nonparametric function-on-function
  regression.
\newblock \emph{Journal of Computational and Graphical Statistics}, {\bf
  31}\penalty0 (3), \penalty0 835--845.

\bibitem[Wilcox(2012)]{Wilcox}
{\rm Wilcox, R.} (2012).
\newblock \emph{Introduction to Robust Estimation and Hypothesis Testing}.
\newblock Elsevier, Waltham, MA.

\bibitem[Yao et~al.(2005)Yao, M{\"u}ller, and Wang]{yao2005}
{\rm Yao, F., M{\"u}ller, H.-G., {\rm and} Wang, J.-L.} (2005).
\newblock Functional linear regression analysis for longitudinal data.
\newblock \emph{The Annals of Statistics}, {\bf 33}\penalty0 (6), \penalty0
  2873--2903.

\bibitem[Yohai and Zamar(1988)]{yohai88}
{\rm Yohai, V.~J. {\rm and} Zamar, R.~H.} (1988).
\newblock High breakdown-point estimates of regression by means of the
  minimization of an efficient scale.
\newblock \emph{Journal of the American Statistical Association: Theory and
  Methods}, {\bf 83}\penalty0 (402), \penalty0 406--413.

\bibitem[Zhou(2021)]{Zhou21}
{\rm Zhou, Z.} (2021).
\newblock Fast implementation of partial least squares for function-on-function
  regression.
\newblock \emph{Journal of Multivariate Analysis}, {\bf 185}, \penalty0 104769.

\end{thebibliography}

\end{document}